\documentclass[preprint,5p]{elsarticle}

\usepackage{lmodern}
\usepackage{amsmath,amssymb,amsfonts}
\usepackage{textcomp}
\usepackage[dvipsnames]{xcolor}
\usepackage{tcolorbox}

\usepackage{csvsimple}
\usepackage{booktabs}
\usepackage{multirow}
\usepackage{ifthen}
\usepackage{colortbl}
\usepackage{tikz}
\usepackage[T1]{fontenc}
\usepackage[font=footnotesize]{caption}
\usepackage{subcaption}
\usepackage{verbatim}
\usepackage{listings}
\usepackage{tablefootnote}
\usepackage{subcaption}

\usepackage[frozencache,cachedir=minted-cache]{minted2}

\usepackage{hyperref}
\usepackage{xurl}

\def\BibTeX{{\rm B\kern-.05em{\sc i\kern-.025em b}\kern-.08em
T\kern-.1667em\lower.7ex\hbox{E}\kern-.125emX}}

\definecolor{box-background}{HTML}{F2F2F2}

\tcbset{
rq/.style={
colback=box-background,
colframe=black,
arc=2mm
}
}

\lstdefinelanguage{JavaScript}{
    morekeywords={
        typeof, new, true, false, catch, function, return, null, switch, var, if, in, while, do, else, case, break, const, let, async, await, throw, try, finally, for, of, export, import, default, class, extends, super, this, yield, continue
    },
    morekeywords=[2]{Array, Boolean, Date, Error, Function, JSON, Math, Number, Object, Promise, RegExp, String, Map, Set, WeakMap, WeakSet, Symbol, console, window, document},
    morecomment=[s]{/*}{*/},
    morestring=[b]',
    morestring=[b]",
    morestring=[b]`,
    morestring=[s][\color{PineGreen}\ttfamily]{/}{/},
    morecomment=[l]//,
    sensitive=true,
}

\lstdefinelanguage{TypeScript}{
    extendedchars=true,
    language=JavaScript,
    morekeywords={
        public, private, protected, readonly, abstract, static, interface, enum, type, implements, keyof, as, satisfies
    },
    morekeywords=[3]{
        string, number, boolean, any, void, never, unknown, null, undefined,
        bigint, symbol, object,
        readonly,
        Awaited, Partial, Required, Readonly, Pick, Omit, Exclude, Extract, NonNullable, Parameters, ConstructorParameters, ReturnType, InstanceType, ThisType, Uppercase, Lowercase, Capitalize, Uncapitalize
    },
    alsoletter={.},
    literate={-}{-}1
}

\usepackage{color}
\definecolor{lightgray}{rgb}{0.95, 0.95, 0.95}
\definecolor{darkgray}{rgb}{0.4, 0.4, 0.4}
\definecolor{editorGray}{rgb}{0.95, 0.95, 0.95}
\definecolor{editorOcher}{rgb}{1, 0.5, 0} %
\definecolor{editorGreen}{rgb}{0, 0.5, 0} %
\definecolor{orange}{rgb}{1,0.45,0.13}		
\definecolor{olive}{rgb}{0.17,0.59,0.20}
\definecolor{brown}{rgb}{0.69,0.31,0.31}
\definecolor{purple}{rgb}{0.38,0.18,0.81}
\definecolor{lightblue}{rgb}{0.1,0.57,0.7}
\definecolor{lightred}{rgb}{1,0.4,0.5}
\usepackage{upquote}
\usepackage{listings}
\lstdefinelanguage{CSS}{
  keywords={color,background-image:,margin,padding,font,weight,display,position,top,left,right,bottom,list,style,border,size,white,space,min,width, transition:, transform:, transition-property, transition-duration, transition-timing-function},	
  sensitive=true,
  morecomment=[l]{//},
  morecomment=[s]{/*}{*/},
  morestring=[b]',
  morestring=[b]",
  alsoletter={:},
  alsodigit={-}
}

\lstdefinelanguage{HTML5}{
  language=html,
  sensitive=true,	
  alsoletter={<>=-},	
  morecomment=[s]{<!-}{-->},
  tag=[s],
  otherkeywords={
  >,
	<!DOCTYPE,
  </html, <html, <head, <title, </title, <style, </style, <link, </head, <meta, />,
	</body, <body,
	</div, <div, </div>, 
	</p, <p, </p>,
	</script, <script,
  <canvas, /canvas>, <svg, <rect, <animateTransform, </rect>, </svg>, <video, <source, <iframe, </iframe>, </video>, <image, </image>, <header, </header, <article, </article
  },
  ndkeywords={
  =,
  charset=, src=, id=, width=, height=, style=, type=, rel=, href=,
  fill=, attributeName=, begin=, dur=, from=, to=, poster=, controls=, x=, y=, repeatCount=, xlink:href=,
  margin:, padding:, background-image:, border:, top:, left:, position:, width:, height:, margin-top:, margin-bottom:, font-size:, line-height:,
  transform:, -moz-transform:, -webkit-transform:,
  animation:, -webkit-animation:,
  transition:,  transition-duration:, transition-property:, transition-timing-function:,
  }
}

\lstdefinestyle{htmlcssjs} {%
  basicstyle={\footnotesize\ttfamily},   
  frame=b,
  xleftmargin={0.75cm},
  numbers=left,
  stepnumber=1,
  firstnumber=1,
  numberfirstline=true,	
  identifierstyle=\color{black},
  keywordstyle=\color{blue}\bfseries,
  ndkeywordstyle=\color{editorGreen}\bfseries,
  stringstyle=\color{editorOcher}\ttfamily,
  commentstyle=\color{brown}\ttfamily,
  language=HTML5,
  alsolanguage=JavaScript,
  alsodigit={.:;},	
  tabsize=2,
  showtabs=false,
  showspaces=false,
  showstringspaces=false,
  extendedchars=true,
  breaklines=true,
  literate=%
  {Ö}{{\"O}}1
  {Ä}{{\"A}}1
  {Ü}{{\"U}}1
  {ß}{{\ss}}1
  {ü}{{\"u}}1
  {ä}{{\"a}}1
  {ö}{{\"o}}1
}
\lstdefinestyle{py} {%
language=python,
literate=%
*{0}{{{\color{lightred}0}}}1
{1}{{{\color{lightred}1}}}1
{2}{{{\color{lightred}2}}}1
{3}{{{\color{lightred}3}}}1
{4}{{{\color{lightred}4}}}1
{5}{{{\color{lightred}5}}}1
{6}{{{\color{lightred}6}}}1
{7}{{{\color{lightred}7}}}1
{8}{{{\color{lightred}8}}}1
{9}{{{\color{lightred}9}}}1,
basicstyle=\footnotesize\ttfamily, %
numbers=left,               %
numbersep=5pt,              %
tabsize=4,                  %
extendedchars=true,         %
breaklines=true,            %
keywordstyle=\color{blue}\bfseries,
frame=b,
commentstyle=\color{brown}\itshape,
stringstyle=\color{editorOcher}\ttfamily, %
showspaces=false,           %
showtabs=false,             %
xleftmargin=17pt,
framexleftmargin=17pt,
framexrightmargin=5pt,
framexbottommargin=4pt,
showstringspaces=false,      %
}%

\lstdefinestyle{json}{
    basicstyle=\ttfamily\small,
    columns=fullflexible,
    showstringspaces=false,
    commentstyle=\color{gray}\upshape,
    morestring=[b]",
    morestring=[b]',
    stringstyle=\color{blue},
    keywordstyle=\color{purple}\bfseries,
    frame=single,
    breaklines=true
}

\setminted{
    linenos,
    fontsize=\footnotesize,
    breaklines,
    numbersep=10pt,
    xleftmargin=13pt,
    frame=single,
}

\hypersetup{
    colorlinks   = true, %
}

\begin{document}
\begin{sloppy}

\title{ModARO: A Modular Approach to Architecture Reconstruction of Distributed Microservice Codebases}

\tnotetext[t1]{This research was joint funded by Adelaide University and Swordfish Computing.}

\author[1]{Oscar Manglaras}%
\ead{oscar.manglaras@adelaide.edu.au}

\affiliation[1]{
    organization={Adelaide University},
    city={Adelaide},
    country={Australia}}

\author[1]{Alex Farkas}%
\ead{alex.m.farkas@gmail.com}

\author[3]{Thomas Woolford}%
\ead{thomas.woolford@swordfish.com.au}
\affiliation[3]{
    organization={Swordfish Computing},
    city={Adelaide},
    country={Australia}
}

\author[4]{Christoph Treude}%
\ead{ctreude@smu.edu.sg}
\affiliation[4]{
    organization={Singapore Management University},
    country={Singapore}
}

\author[5]{Markus Wagner}%
\ead{markus.wagner@monash.edu}
\affiliation[5]{
    organization={Monash University},
    city={Clayton},
    country={Australia}
}

\begin{abstract}
Microservice architectures promote small, independently developed services,
but increase overall architectural complexity.
It is crucial that developers understand the architecture and how
changes to a service affect the overall system,
but rapid and independent development of services
increases the risk of architectural drift and discourages the
creation and maintenance of documentation.
Automatic architecture reconstruction
can help avoid these issues, but
it is difficult to reuse reconstruction code across multiple projects,
as all use different combinations
of technologies and project-specific conventions.
Reconstruction of architecture-level details is further complicated by the tendency to split microservices into
separate repositories, preventing a full view of the system from any one codebase.
In this paper, we present and evaluate ModARO, an approach to microservice architecture reconstruction
that allows writing modular reconstruction code (`extractors') for any technologies and reusing
them across different projects, independent of the surrounding technology
stack or whether or not the services are split into multiple codebases.
We demonstrate the effectiveness of our approach by configuring ModARO to
reconstruct 10 open source projects,
and we validate the usefulness and usability of ModARO against
a state-of-the-art baseline in
a user study with 8 industry practitioners.
Using this approach, developers can assemble or create
extractors tailored to their technology
stacks and distribute architecture reconstruction across repositories,
enabling integration into repository CI/CD pipelines.
\end{abstract}


\maketitle

\section{Introduction}

A microservice architecture is a type of
distributed software architecture in which a system is composed
of multiple small, independent components,
called `microservices'.
Each microservice is a standalone, fully deployable piece of software dedicated to
a single business capability, and can be developed independently by different developers~\cite{Lewis2014}.
While this approach keeps individual services relatively simple,
the large number of services and the intricate communication behaviour between services make the
overall architecture of the system increasingly complex~\cite{Kleehaus2019}.
Therefore, it is crucial that developers understand the architecture
and maintain visibility of how changes to any individual microservice can affect the overall system,
and vice versa.
Microservice architectures are closely associated with the principles of
`continuous software development',
an agile approach with a focus on using tooling
to automate deployment and testing,
enabling more rapid development
and release of software~\cite{OConnor2017,Theunissen2022}.
However, these practices increase the risk of architectural drift,
while at the same time
de-incentivising the creation and maintenance of documentation,
which can lead to documentation becoming
out of sync with the actual architecture of
the system~\cite{Islam2023,Kleehaus2019}.

Architecture reconstruction, the process of
extracting the architectural information of a system from artifacts such as source code and
configuration files~\cite{Cerny2022a}, can help address these issues~\cite{Theunissen2022}.
The detected architecture can be used to generate documentation,
perform architectural conformance checks to avoid architectural drift,
and flag the impacts of any changes made between commits.
These tasks can be automated using the
continuous integration and deployment (CI/CD)
pipelines of microservice repositories,
ensuring that checks are performed on every version of the code and
that documentation is always up-to-date.
Generating documentation in this way is called ``continuous documentation''~\cite{Andel2022}.

However, a general approach to microservice architecture reconstruction of
static artifacts is challenging.
Microservices use different combinations of technologies (programming languages,
libraries, deployment technologies, etc.) and project-specific conventions; these
technological differences can be seen even between microservices in the same
architecture~\cite{AmorosodAragona2024}, making it difficult to reuse reconstruction
code between projects.
Architecture reconstruction is further complicated by the tendency to split
microservices into separate repositories (multi-repo projects), which means that reconstruction code
would only have access to a single microservice if run in its repository.

To address these challenges, we present
ModARO\footnote{\url{https://gitlab.com/swordfish-computing-group/modaro}}
(MODular Architecture Reconstruction Orchestrator),
an approach to static architecture reconstruction
that allows writing modular reconstruction code (`extractors') for any technology and
reusing them across different projects. The core of our approach is the use of a shared architecture model
to pass information between extractors and an algorithm that orchestrates when extractors run;
together, these allow extractors to be written for individual technologies and reused
between microservices, without concern for what other
technologies are in use.
This differs from existing approaches, which only work with specific technologies,
only work with code files (as opposed to configuration files, build scripts, etc.),
or require manual intervention to reuse reconstruction code between projects.
We also designed ModARO so that each microservice repository
in a multi-repo project can be reconstructed individually, and the output combined into
a single architecture separately when and where the microservices are collected for deployment;
we call this concept `distributed architecture reconstruction', and are not aware of any
existing approaches that support this workflow.

In this paper, we explain the key design features of ModARO and
present a case study in which we extract architectural information
documented by ten
open source microservice projects
to validate that our approach is capable of reconstructing a range of
microservice projects and that we achieved our design aims.
We have not identified any static artifacts that ModARO is unable to extract,
and we found that our approach successfully
facilitates modularity and supports architecture reconstruction in distributed codebases.
Additionally, we conduct a user study involving 8 industry
practitioners to gather feedback on
ModARO and compare it with a state-of-the-art baseline (ReSSA~\cite{Schiewe2022a}),
which found that ModARO was more useful and easier to use.

\section{The ModARO Approach}
The goal of our approach is to allow static architecture reconstruction
code to be reusable across diverse microservice projects.
To achieve this goal, we have two primary design aims:
modularity and support for distributed codebases.

\subsection{Modularity}
Extracting architectural information from static files is
often relatively simple; the information may be stored in configuration
files that can be read with a parsing library (e.g. Docker Compose YAML files)
or represented in code that can be detected with
string or regex searches (e.g. Java Spring\footnote{\url{https://spring.io/}}
annotations or particular function calls). However,
each microservice project (and each microservice within each project) can
use unique combinations of these technologies, which prevents the same
reconstruction code from being easily reused. The task is further complicated
by the frequent need to use the information collected from one technology to inform
the extraction of data from another technology, e.g. using the build
directories stored in a Docker Compose file to identify which microservice
is calling a particular function in a repository containing
multiple microservice codebases (mono-repo projects).
To address this issue, our approach aims to break reconstruction code
into reusable and interoperable modules that can be assembled to address unique project
technology stacks; we call these modules `extractors'.

\subsubsection{Architecture Model}
Extractors are made interoperable by interfacing exclusively through a shared
model of the microservice architecture, which we refer to as simply
`the model' throughout this paper.
We refer to architectural entities within the model (e.g., microservices,
HTTP requests, etc.) as `model entities`.
When extractors gather architectural information,
they write the results into the model,
where other extractors can read them.
For example, the model may contain an array of microservice entities,
each conforming to a specific ``microservice entity schema'', which specifies
that microservice entities have a \texttt{\$path} field indicating the directory of the
code of that microservice. An extractor that detects microservices
would add new microservice entities to the model and populate
their \texttt{\$path} fields in conformance with the schema.
Other extractors, trusting that data is stored according to the agreed schema, could then use 
these paths to analyse the source code of the microservice and place
their results into the corresponding microservice entities in the model.
Using an architecture model as an interface
means that the individual extractors providing a piece of information
can be swapped without affecting the functioning of the other extractors
(as long as all agree on the schema of the architecture model).
For example, an extractor that creates microservice entities and sets
their \texttt{\$path} fields by reading Docker Compose files could be
swapped for one that does so by finding directories that contain
\texttt{pom.xml} files, and it would not affect any extractors using
the \texttt{\$path} field to find HTTP requests.

The model is recursive, allowing model entities to be stored within each other;
the top-level of the model is a model entity containing information
about the overall architecture/project,
within which are model entities containing information about, e.g., microservices.
The final version of the model, after any extractor modifications, is what is output at
the end of our approach (Figure~\ref{fig:model/in-out} shows an example of this).
Note that the development of a comprehensive model schema that can
represent arbitrary microservice architectures is beyond
the scope of this paper and would be a topic of future work.
A schema developed to test our approach is visible throughout this paper, but we
make no claims about its wider applicability, only that it is sufficient to demonstrate and
validate our approach.
Indeed, we expect to need to store unanticipated types of information,
and so require that the model allow setting arbitrary fields for model entities.
If a more comprehensive schema is developed in the future,
such dynamic adaptation may not be necessary.

The above behaviour requires that the model storage and transmission format allow
arbitrary key-value fields that can be recursively stored inside each other.
JSON met these requirements, and is how we present model examples in this paper.
Extractors must be able to identify the type of architectural concept that a particular
model entity is representing; to do this,
we gave each model entity a \texttt{\$TYPE} field indicating the
kind of entity (for example, \texttt{\$TYPE: "microservice"} for a microservice entity).
Extractors may also need to share information
that should not appear in the final output (e.g., the local directory containing a microservice).
The model must hence support setting these (potentially arbitrary) fields,
which will be removed before the final model is output.
In our implementation, we did this by reserving keys that match a specific pattern
(\texttt{\textasciicircum\textbackslash\$[a-z0-9\_]+\$}~\footnote{\url{https://regex101.com/r/UoFYHU/1}}),
which are stripped from the output.

\subsubsection{Extractors}
\begin{figure}
    \centering
\begin{minted}{TypeScript}
import { getPaths, register_extractor } from "api";

async function extractor(entity) {
  const glob = entity.$path + "/**/*.java";
  const javaFiles = await getPaths(glob);
  if (javaFiles.length > 0)
    entity.java = true;
  return entity;
}

register_extractor(extractor, {
    properties: {
        $TYPE: { const: 'microservice' },
        $path: { type: 'string' },
    },
    required: ['$TYPE', '$path'],
})
\end{minted}
    \caption{%
        A trivial example of an extractor that run on microservice entities using our
        JavaScript implementation.
        It tests if the microservice source code contains any Java files,
        and if so it sets \texttt{java} field in the model entity to \texttt{true}.
    }
    \label{fig:extractor-fn}
\end{figure}
\begin{figure}[t]
    \centering
    \includegraphics[width=\linewidth]{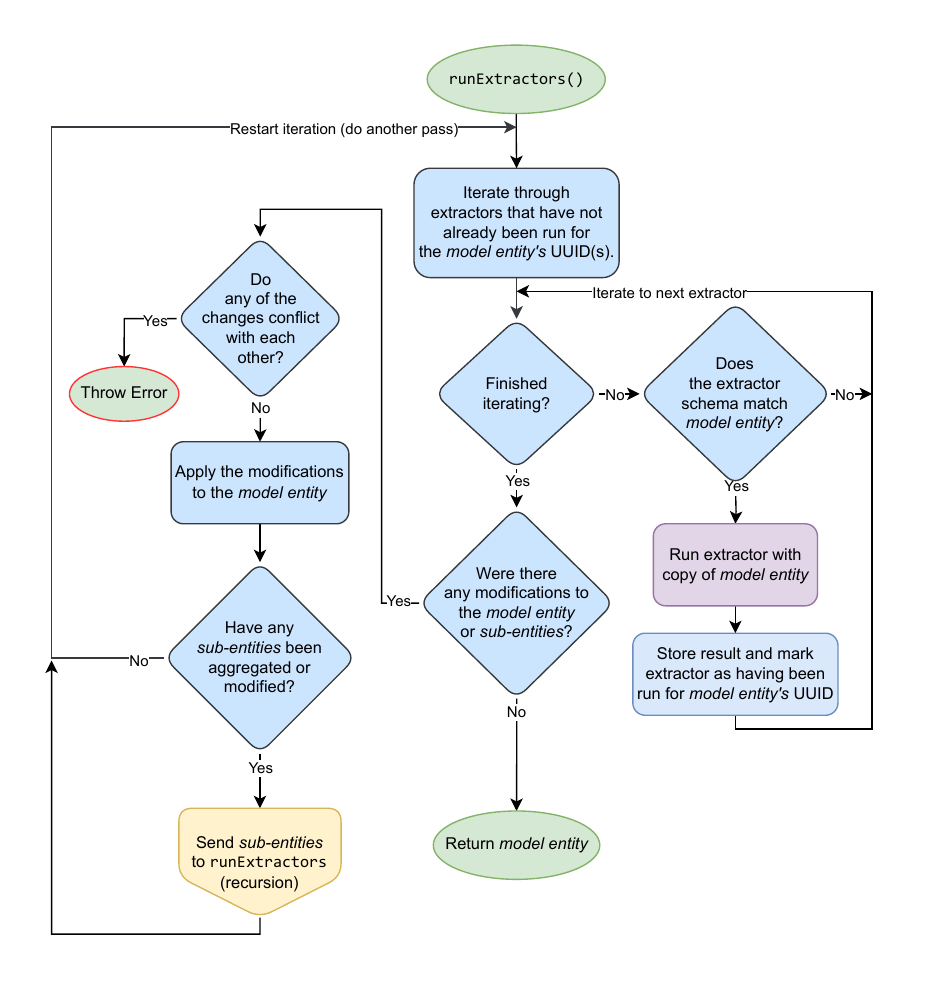}
    \caption{
        The reconstruction algorithm. This is run on
        each model entity.
    }
    \label{fig:run-extractors-simplified}
\end{figure}

Extractors are modules of code that accept a model entity
as input, modify the model entity to store extracted information,
and return the modified entity as output.
Our implementation uses JavaScript functions for each extractor,
with the model entities represented by normal JavaScript objects
(associative arrays) that are passed into the function
as an argument.
Figure~\ref{fig:extractor-fn} shows an example of a very simple extractor
that sets the \texttt{java} field on the model entity to \texttt{true}
if the directory specified by the \texttt{\$path} field contains any Java files.
The extractor function must then be registered with the reconstruction algorithm.
In our implementation, this was done by passing the function to a 
\texttt{register\_extractor} API method.
We intend for extractors to have no global or persistent state; instead, the
behaviour of the extractor is defined entirely by the contents of the input model
entity.
In practice, the behaviour of the extractor will also
depend on the contents of the code repository being analysed; however, we assume that
the repository is not being modified during the reconstruction process.
The stateless design would theoretically
allow extractors to be executed in parallel or memoised, improving performance.
We also intend for extractors to have no restrictions on their scripting capabilities.
They may load third-party libraries and invoke external processes,
allowing extractors to act as a wrapper around existing tools to
integrate them into our approach.
We acknowledge that this level of flexibility introduces potential security
risks, particularly when running extractors from third-party sources.
While these concerns fall outside the scope of this paper,
they may be worth exploring in future work.

We expect that extractors will want to operate on specific types of model
entities (e.g., only operate on microservice entities) or require specific
fields to be present before doing so (e.g., that a \texttt{\$path} field has been
set by another extractor). 
To support this behaviour, extractors must register a schema
that specifies the structure of the model entities they accept
as input. We used the JSON Schema
specification\footnote{\url{https://json-schema.org/}} for this
purpose in our implementation.
Figure~\ref{fig:extractor-fn} shows such a schema being registered
alongside the extractor function, specifying that any model entities passed to the extractor must
have the \texttt{\$TYPE} field set to \texttt{microservice} (ensuring that it
is passed a microservice entity), and have the
\texttt{\$path} field set to a string. This ensures that all fields of the
microservice entity that the extractor function uses are populated before it is run.

Our intention is that extractors would be provided with an API for common
operations, such as file searches and file parsing. Figure~\ref{fig:extractor-fn}
shows an example of this with the \texttt{getPaths} function, which returns a list of
all files that match the given glob pattern.
We provided extractors with a basic set of API functions for file searching and
regular expression searches during our studies,
but we would expect these to be expanded and improved if our approach were
widely adopted.

\subsubsection{Reconstruction Algorithm}
\begin{figure}[!t]
    \centering
    \begin{subfigure}{\linewidth}
\begin{minted}{javascript}
async function extractor(model) {
  model.microservices = [];
  model.microservices.push(
    await createModelEntity('microservice', {
      $path: `${model.$path}/service1`,
      name: 'service1',
    })
  );
  return model;
}
registerExtractor(extractor, {
  type: 'object',
  properties: {
    $TYPE: { const: '$MODEL' },
    $path: { type: 'string' },
  },
  required: ['$TYPE', '$path'],
});
\end{minted}
        \caption{
            An extractor that specifies that its input must have the
            \texttt{\$TYPE} field set to \texttt{\$MODEL} (which we use for
            the top-level model entity) and the \texttt{\$path} field set to a string
            to denote the local file path of the code repository.
            It creates a new microservice called \textit{service1} and pushes
            it to the \texttt{microservices} array.
        }
        \label{fig:extractors/in-out/model}
    \end{subfigure}
    \vspace{0.5em}
    
    \begin{subfigure}{\linewidth}
\begin{minted}{javascript}
async function extractor(microservice) {
  const path = microservice.$path;
  const glob = `${globEscape(path)}/**/*.java`;
  const javaFiles = await getPaths(glob);
  if (javaFiles.length > 0) {
    microservice.java = true;
  }
  return microservice;
}

registerExtractor(extractor, {
  type: 'object',
  properties: {
    $TYPE: { const: 'microservice' },
    $path: { type: 'string' },
  },
  required: ['$TYPE', '$path'],
});
\end{minted}
        \caption{
            An extractor that specifies that its input must have the
            \texttt{\$TYPE} field set to \texttt{microservice} (which we use for
            microservice entities) and the \texttt{\$path} field set to a string
            to denote the local file path of the microservice code directory.
            The microservice entity created in Figure~\ref{fig:extractors/in-out/model}
            conforms to these requirements, and so is passed to this extractor,
            which performs the same operation as Figure~\ref{fig:extractor-fn}.
        }
        \label{fig:extractors/in-out/microservice}
    \end{subfigure}
    \caption{
        An example of two extractors that could run on the same project.
        The first creates the model entity that triggers the second.
        Figure~\ref{fig:model/in-out} shows the contents of the model
        before and after being passed to the extractors.
        }
    \label{fig:extractors/in-out}
\end{figure}

\begin{figure}
    \centering
    \begin{subfigure}{\linewidth}
\begin{minted}{json}
{
  "$TYPE": "$MODEL",
  "$VERSION": "0.0.1",
  "$path": "/path/to/repo"
}
\end{minted}
        \caption{The initial model passed to the algorithm.}
    \end{subfigure}
    \vspace{0.5em}

    \begin{subfigure}{\linewidth}
\begin{minted}{json}
{
  "$TYPE": "$MODEL",
  "$VERSION": "0.0.1",
  "$path": "/path/to/repo",
  "microservices": [
    {
      "$TYPE": "microservice",
      "$path": "/path/to/repo/service1",
      "name": "service1",
      "java": true
    }
  ]
}
\end{minted}
        \caption{
            The final version of the model output by the algorithm (assuming that
            \textit{service1} did indeed have Java files). Normally the \texttt{\$path} fields
            would be removed as the local repository path would be unique
            to the specific machine that ran the reconstruction.
        }
    \end{subfigure}
    \caption{The initial and final form of the model passed to the extractors in Figure~\ref{fig:extractors/in-out}.}
    \label{fig:model/in-out}
\end{figure}

The reconstruction algorithm defines the execution flow of extractors and the process
of integrating their outputs into the model, detecting any conflicts.
The algorithm is triggered recursively; first, we create the top-level
model entity with initial values
(e.g. the file path of the repository being analysed).
Then we pass this top-level model entity to the algorithm
(Figure~\ref{fig:run-extractors-simplified}), which forwards it to
all extractors whose registered schema indicates that they accept top-level
entities. To denote the top-level entity, we set the
\texttt{\$TYPE} field to \texttt{\$MODEL}, as shown in
Figure~\ref{fig:model/in-out}.
These top-level extractors collect and store architecture-level information,
which includes creating new microservice entities, which triggers the algorithm
to run microservice-specific extractors on these microservice entities.
This recursive process continues
until no new entities are created or there are no more extractors with
conformant schemas to run.
After a model entity has been modified, the algorithm does another
pass over the list of extractors in case the modification caused conformance
to another extractor schema.
A simple example of this is shown in Figures~\ref{fig:extractors/in-out}
and~\ref{fig:model/in-out}.

In this way, the algorithm allows extractors to implicitly pass data to
each other by writing and reading fields within the model.
When an extractor requires a specific piece of information, it will only be run
after that information has been collected and stored in the model by another
extractor; extractors do not need to know which other extractor is providing
the information, nor do they or the user of ModARO need to manually specify
the order in which extractors should be run. This allows extractors to be
reused between projects without code changes or manual interventions.

\subsection{Support for Distributed Codebases}
\begin{figure}[!t]
    \centering
    \begin{subfigure}{\linewidth}
        \begin{minted}{json}
{
  "flags": {"flag1": true},
  "microservices": [
    {
      "name": "service1",
      "java": true
    },
    {
      "name": "service2",
      "java": false
    }
  ]
}
        \end{minted}
        \caption{Input 1}
    \end{subfigure}
    
    \begin{subfigure}\linewidth
        \begin{minted}{json}
{
  "flags": {"flag2": true},
  "microservices": [
    {
      "name": "service1",
      "version": "1.0.2"
    },
    {
      "name": "service3",
      "version": "2.3.4"
    }
  ]
}
        \end{minted}
        \caption{Input 2}
    \end{subfigure}
    
    \begin{subfigure}{\linewidth}
        \begin{minted}{json}
{
  "flags": {"flag1": true, "flag2": true},
  "microservices": [
    {
      "name": "service1",
      "java": true,
      "version": "1.0.2"
    },
    {
      "name": "service2",
      "java": false
    },
    {
      "name": "service3",
      "version": "2.3.4"
    }
  ]
}
        \end{minted}
        \caption{
            Output.
            The two versions of \textit{service1} have been combined into single entry
            containing the unique information from both versions.
        }
    \end{subfigure}
    \caption{An example of model aggregation. Non-relevant fields like \texttt{\$TYPE} have been
    removed for brevity.}
    \label{fig:model-aggregation}
\end{figure}
\begin{figure*}
    \centering
        \begin{subfigure}{.32\linewidth}
        \includegraphics[width=\linewidth]{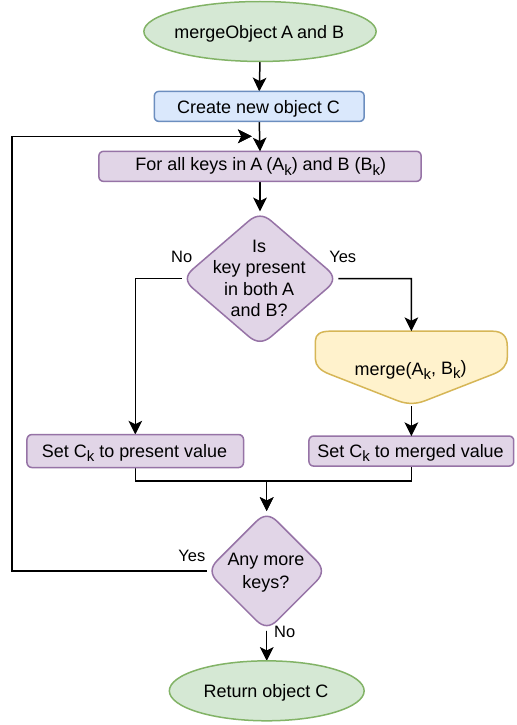}
        \caption{Aggregating two objects.}
        \label{fig:aggregation-algorithm/object}
    \end{subfigure}
    \hfill
    \begin{subfigure}{.4\linewidth}
        \includegraphics[width=\linewidth]{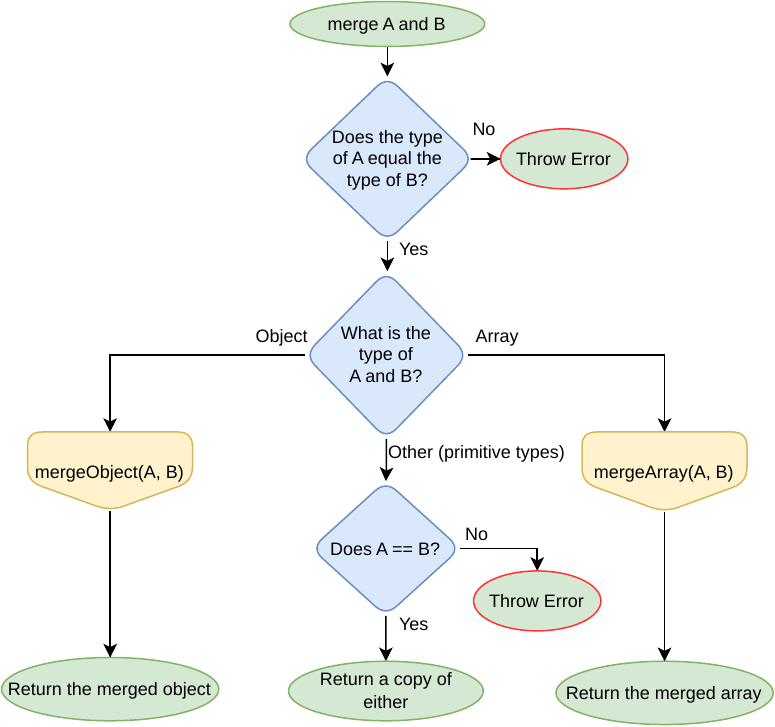}
        \caption{
            Aggregating two fields of unknown types. Conflicts throw errors.
            In practice one may wish to catch these errors in
            Figure~\ref{fig:aggregation-algorithm/object} and print warnings
            to avoid terminating the entire aggregation process.
        }
        \label{fig:aggregation-algorithm/merge}
    \end{subfigure}
    \hfill
    \begin{subfigure}{.2\linewidth}
        \includegraphics[width=\linewidth]{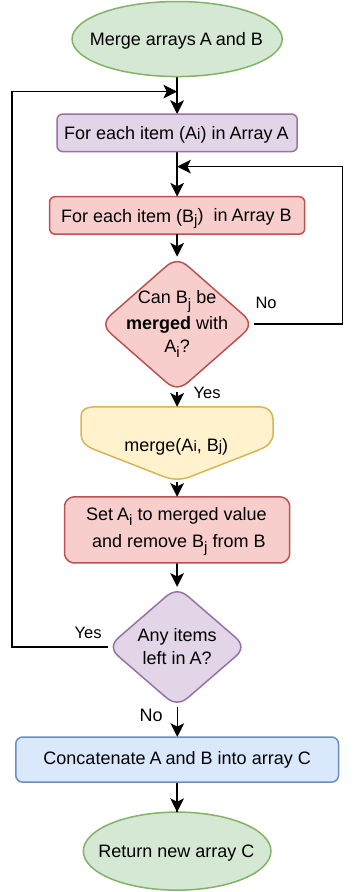}
        \caption{
            Aggregating two arrays.
        }
        \label{fig:aggregation-algorithm/array}
    \end{subfigure}
    \caption{The aggregation algorithm used to combine multiple models.}
    \label{fig:aggregation-algorithm}
\end{figure*}

Microservice projects can span
multiple repositories; microservices may evolve independently or be versioned
separately, and the same microservice may be used in multiple such
projects. This poses difficulties for reconstruction approaches that assume
access to all services during analysis.
While it is possible to pull all repositories and reconstruct
the architecture centrally, doing so negates many of the benefits of a multi-repo setup;
even minor changes to one service would require re-analysing the
entire system, and reconstruction could not be integrated into a
single service's CI/CD pipeline without knowing which other
services will be deployed alongside it.
This undermines the loose-coupling principle that is a common
characteristic of how microservice architectures are developed and
deployed~\cite{Lewis2014}, and fails to support their
compositional nature, where a service may serve as a reusable
component in multiple systems, as is the case with our industry partner
Swordfish Computing.
Our approach aims to address this challenge by allowing
individual microservice repositories to be reconstructed separately,
and their outputs collected and aggregated at the same time as the
microservices themselves.
This is an approach that
we have dubbed \emph{distributed architecture reconstruction}.

\subsubsection{Model Aggregation}

Our aggregation algorithm defines how the outputs of multiple
reconstruction operations are combined into a single model
of the entire architecture.
It may seem that each independent reconstruction run targets a
single microservice repository and that its output fully specifies that microservice.
However, in practice, relevant microservice data can also reside outside
the codebase, for example in configuration files stored elsewhere, such
as repositories containing the deployment configuration for the full microservice system.
Examples of this behaviour
can be seen in mono-repo projects as well, where relevant information
can be stored in deployment files (e.g., Docker Compose) in the root of the
repository, outside of any of the microservices' subdirectories.
As such, we must assume that information about any particular
model entity may be distributed across multiple repositories
and that there may be multiple instances of the same underlying
entity that must be aggregated together.

Figure~\ref{fig:aggregation-algorithm} presents our algorithm
that fulfills these requirements. It operates like a recursive union,
merging the fields of several model entities into one.
Conflicting fields throw errors,
which guarantees that the user is aware that different reconstruction
operations are producing conflicting results. Some form of automated conflict
resolution could be possible, but it was deemed out of scope for our research.
Figure~\ref{fig:model-aggregation} demonstrates this behaviour with the
\texttt{flags} field.
The algorithm also allows aggregating arrays; 
when aggregating two arrays, we iterate through them and see if any array
elements can be themselves aggregated together; those that can are, and those that
cannot are appended. This behaviour is demonstrated in Figure~\ref{fig:model-aggregation},
which shows the result of aggregating two arrays of microservices.
In this example, we can see that \emph{service1} was present in both inputs,
but different information about the service was collected, all of which are present
in the output.
This demonstrates how our approach can be used to collect information about
the same microservice from multiple sources.

This aggregation algorithm is also used by the reconstruction algorithm
to combine the outputs of multiple extractors that modify the same
model entities. This means that an extractor that collects information from deployment
configuration files operates the same in both mono- and multi-repo projects;
in both cases, the extractor can create new microservice entities to hold
the deployment information, and in both cases that entity will be aggregated
with instances of the same entity created by other extractors.

\subsubsection{Retroactive Linking}
\label{section:design/links}
\begin{figure}
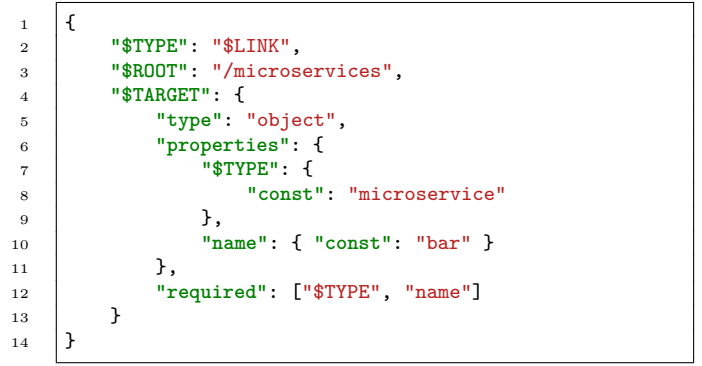

    \centering
    \begin{minted}{json}
{
    "$TYPE": "$LINK",
    "$ROOT": "/microservices",
    "$TARGET": {
        "type": "object",
        "properties": {
            "$TYPE": {
                "const": "microservice"
            },
            "name": { "const": "bar" }
        },
        "required": ["$TYPE", "name"]
    }
}
    \end{minted}
    \caption{A link targeting a microservice with the name \texttt{bar}.}
    \label{fig:link}
\end{figure}

When extracting the architectural details of an individual
microservice, we sometimes need to specify a relationship to
another service. For example, we may want to specify the dependencies between
microservices, or that one service is making an HTTP request to another
specific service. To support such relationships, the model needs
some way of representing another service (e.g., a unique string ID);
in this paper, we will refer to these
representations as ``links'', as they effectively act as references to
a different model entity.
However, distributed architecture reconstruction makes
links more challenging to specify, as we cannot assume what knowledge extractors
will have about the remote service being linked.
We cannot rely on extractors knowing the URI of
the remote service, nor on unique ID values being available.

To address these challenges, we present a concept called
retroactive linking; defining a link to another model entity in the
model which is only resolved (retroactively) at a later time
(i.e., after aggregation).
We propose that the link must be able to describe a schema for its target,
e.g., specifying the data type of specific fields, or that fields must
equal specific values or conform to restraints (numbers within specified ranges,
strings that match a given regular expression, etc.).
After aggregation, we can then search the model for conformant
model entities and resolve links to them.
The advantage of using a schema is that it allows links to be defined
using whatever arbitrary information is available about the link target.
For example, if a remote microservice is only referenced by its domain name,
then that domain name could be used to define the link target.
This approach also helps keep the aggregated models accurate in the
face of asynchronous code changes; e.g.,
if the remote service changed its domain name, then the
link would automatically fail to resolve, without any intervention needed
by the service that defines the link. Such link failures could be
used as an indicator of errors in the microservice architecture, and
comparing changes in link resolution could be used diagnostically when
updating microservices.
In our implementation and in any examples that we show, we use JSON schema
to describe the link targets (see Figure~\ref{fig:link}).

\section{Research Questions and Studies} \label{section:research-questions}

To evaluate our approach, we ran a case study in which we
reconstructed
the architectures of ten open source microservice projects on Github. 
By testing our approach against projects with different programming languages,
different libraries, and
different code conventions, we can verify that our approach
is capable of reconstructing a range of microservice architectures.
\begin{enumerate}
    \item [RQ1] To what extent can our approach reconstruct architectural information from diverse microservice codebases?
\end{enumerate}
To answer this question, we manually identified documented
architectural information in the ten projects,
implemented extractors for each piece of information,
and recorded if any information could not be extracted and why.
We also want to investigate whether our approach
achieves our design aims.
\begin{enumerate}
    \item [RQ2] How reusable are extractors across different
        microservice projects and what factors limit their reuse?
    \item [RQ3] How well does our approach support distributed
        architecture reconstruction across multiple microservice
        repositories?
\end{enumerate}
To answer RQ2, we manually inspected the 71 extractors created as
a result of RQ1, classified them, and analysed the reasons why
some extractors were not reusable.
To answer RQ3, we tracked whether any extractors needed to
be modified to support distributed reconstruction and whether any information
could not be collected due to the distributed codebase.

The case study validates capabilities, but
we also want to evaluate the usefulness and usability of ModARO in-practice;
no matter how capable our approach is, it has little consequence
if the user experience is so hostile that people are deterred from using it.
Our last research question investigates these concerns.
\begin{enumerate}
    \item [RQ4] How do practitioners perceive the usefulness
    and usability of our approach, and what factors contribute to or could enhance these aspects?
\end{enumerate}
To answer this question, we conducted a study with 8 industry practitioners who manually
performed tasks using ModARO and a state-of-the-art baseline (ReSSA~\cite{Schiewe2022a}),
and then provided
structured quantitative and qualitative feedback on their experience.

\section{Github Case Study} \label{section:github-study}
\begin{table*}[t]
    \centering
    \caption{
        The 10 study projects.
        The service count is based on the project documentation and does
        not include databases or message brokers.
    }
    \resizebox{\linewidth}{!}{%
    \begin{tabular}{ll rrlc}
    \toprule
        ID&name &   LOC  &\# services&Github repo & multi-repo  \\
    \midrule
        P01&StaffJoy V2&  232,972  &14&LandRover/StaffjoyV2  \\
        P02&Cloud Native Strangler Example&  105,157  &10&kbastani/cloud-native-microservice-strangler-example  \\
        P03&E-Commerce App&  20,395  &5&venkataravuri/e-commerce-microservices-sample  \\
        P04&Train Ticket&  343,167  &41&FudanSELab/train-ticket/  \\
        P05&Hipster Shop&  29,141  &11&GoogleCloudPlatform/microservices-demo  \\
        P06&Piggy Metrics&  19,942  &9&sqshq/PiggyMetrics  \\
        P07&Photo uploader (NGINX Fabric Model)&  30,020  &7&nginxinc/mra-ingenious  \\
        P08&eShop on Containers:&  137,029  &15&dotnet-architecture/eShopOnContainers  \\
        P09& Spinnaker&   1,143,296 &12&spinnaker/spinnaker & yes\\
        P10& Lelylan&  113,562 &15&lelylan/lelylan & yes\\
    \bottomrule
    \end{tabular}
    }
    \label{tab:applications}
\end{table*}

\subsection{Study Design}
For the case study, we did not want to define the specific architectural details
to extract from each project, as we thought this would bias our evaluation;
not all projects may want or need to document the same information.
To address this concern, we instead attempted to extract only the information
already presented in the existing documentation for each project:
the information the developers thought was important enough
to document. This also had the advantage of forcing us to attempt extraction of
a wider range of information, which we would have been unlikely to select ourselves.

Projects were chosen from the
Microservice Dataset, Extended Version\footnote{\url{https://github.com/davidetaibi/Microservices_Project_List/tree/ed84473c832c4f4c6f9c828c415175c79eac42d6}}~\cite{Rahman2019}.
We selected eight ``demo/toy projects'' and two
``Industrial or Production-Ready Projects''.
We filtered the dataset according to our exclusion criteria:
\begin{itemize}
    \item [E1] Project lacks any dedicated documentation.
    \item [E2] Code and/or documentation not in English.
    \item [E3] Documentation does not specify any architectural details, only build/deployment instructions.
    \item [E4] Code no longer available.
    \item [E5] Application superseded by another application in the dataset.
    \item [E6] Code not hosted on Github.
    \item [E7] Used in pilot studies.
\end{itemize}
And to exclude the more trivial
demo projects, we sorted the projects according to their lines of code (LOC) with the
\texttt{cloc}\footnote{\url{https://github.com/AlDanial/cloc}} tool.
We then selected 8 demo projects at random from the top 20 and selected two of the production-ready
projects. The 10 projects used in our study are shown in Table~\ref{tab:applications}.
For each project, we performed the following steps:
\begin{enumerate}
    \item Identify each distinct type of information in the documentation
        (microservice names, HTTP endpoints, programming languages, etc.).
        The presentation format (graph, table, prose, etc.) is ignored.
    \item Manually analyse the codebase and identify how or if each
        type of information is encoded.
    \item If possible, reuse or create one or more extractors to collect
        each piece of information.
    \item Compare the output of the extractors with the documented information
        and attempt to fix any discrepancies. For larger applications such
        as Train-Ticket and Spinnaker, we lacked time to verify all information,
        so we checked a random sample (minimum 8) of the output
        (e.g., if an extractor identifies the HTTP endpoints of a microservice,
        we randomly sample microservices and compare a random sample of documented
        endpoints to our output).
\end{enumerate}
The data we collected about the projects and
extractors\footnote{\url{https://doi.org/10.25909/31237213}}
and the code for those extractors\footnote{\url{https://github.com/oscar-manglaras/ModARO-Github-Study-Extractors}} can be found online.

An important aspect of ModARO is that extractors are provided APIs to
assist with common tasks. While they are not under investigation
in this paper, we still provided an initial set of APIs to reduce code duplication,
and hence the time spent writing each extractor.
Specifically, we provided APIs for file searching, regular expression
matching, and exposed existing parsers for JSON, XML, YAML, and TOML.
To make regular expressions less arduous to write and easier to
read, we provided extractors with
a library called Super-Expressive\footnote{\url{https://github.com/francisrstokes/super-expressive}},
and we also provided pre-made expressions for some common patterns (URIs, string literals, etc).

\subsection{\textbf{RQ1 - To what extent can our approach reconstruct architectural information from diverse microservice codebases?}}
\label{section:rq1}

\begin{table*}
    \centering
    \includegraphics[width=\linewidth]{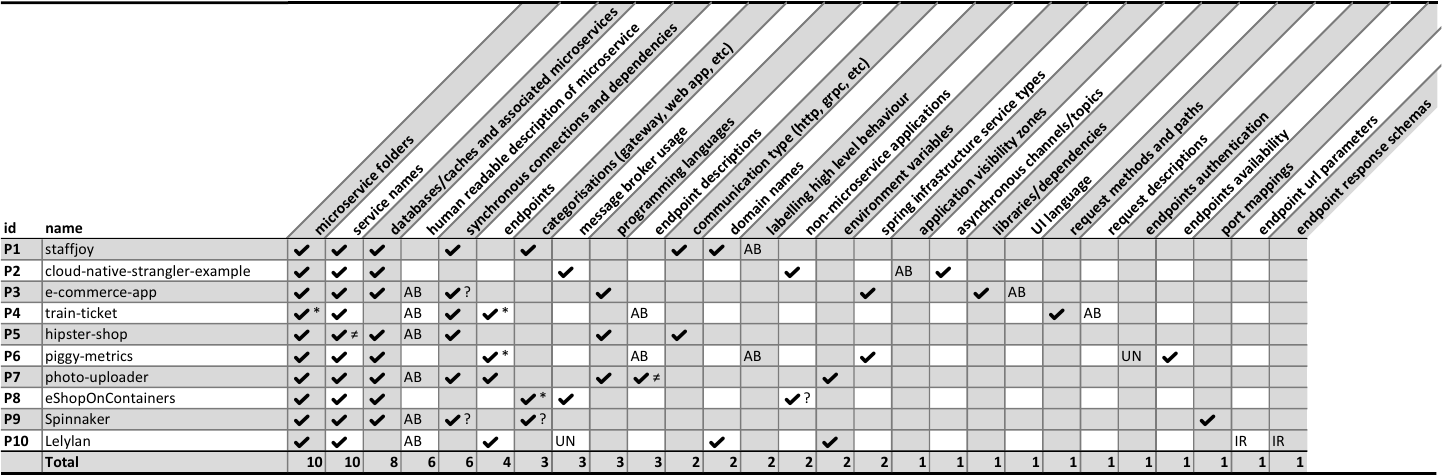}
    \caption{
        Type of documented information collected for each application/project.
        Blank cells for a project indicate that the information was not documented,
        checkmarks indicate that we were able to extract the information,
        and the two-letter codes denote information that was documented,
        but that we were unable to extract. 
        Symbols after the checkmarks denote discrepancies between our output and the
        documentation.
    }
    \label{fig:documentation-data}
    \label{table:documentation-data}
\end{table*}

In this section, we validate that ModARO is capable of extracting a
range of architectural information
from a diverse set of technologies. To do so, we catalogue the architectural
information that our extractors were or were not able to collect during the
study, and the programming languages, frameworks, and other technologies
used by those projects.
Note that this is not intended to be a definitive survey of the projects; there may be
documented information that we missed and technologies that we did not recognise.
What is important is that the range of information and technologies that we identified is
extensive and diverse enough to answer the research question.

\subsubsection{Documented Information}

Table~\ref{fig:documentation-data} shows the types of architectural information
documented in the ten projects and whether our extractors were
able to collect them successfully.
Certain kinds of data were especially common, such as the names of services,
the presence and use of databases or caches, and the connections between services
via requests or dependencies. These aspects were often visualised using
network graphs or adjacency matrices, and reflect the types of architectural
features most commonly targeted by existing reconstruction research
\cite{Granchelli2017,Engel2018,Mayer2018,Kleehaus2018,Alshuqayran2020,Bushong2021}.

However, Table~\ref{fig:documentation-data} also shows that
there are types of information that are less consistently
documented.
These include
service categorisations, programming languages and libraries,
endpoint URL parameters, and communication protocols.
While some of these details have been addressed in prior work, no single
approach has attempted to collect the full breadth of this information.
This is unsurprising as existing reconstruction methods tend to focus
on common architectural elements which they can assume are present in
most or all microservice projects.
In contrast, many of the details observed here are either
highly project-specific (e.g., custom microservice categorisations,
important environment variables, etc.) or lack standard
encoding conventions (e.g., front-end technologies or protocol use),
making them difficult to extract using generic tools.
The prevalence of such varied documentation data serves as an endorsement
of our approach. By structuring analysis around lightweight, modular extractors,
developers can tailor data collection to the needs of their specific
project and are not limited to just the most common architectural details.

\subsubsection{Documentation data that was not reconstructed} \label{section:doc-not-reconstructed}

Some information from the documentation was not collectable by our extractors, indicated
by the two-letter codes in Table~\ref{fig:documentation-data}.
These fall into the following categories:
\begin{enumerate}
    \item[\textbf{AB}] \textbf{Abstract:} 
    Information that is not present in the repository due to the data
    being too high-level or abstract.
    For example, written descriptions of microservices.

    \item[\textbf{IR}] \textbf{Insufficient Resources:} Information that was theoretically present
    in the code, but for which we lacked the time and resources to attempt to extract.

    \item[\textbf{UN}] \textbf{Unknown:} Information for which we were unable to identify
    how or if the data was encoded in the repository. The result of either
    inaccurate or unclear documentation, or our inexperience with the source
    code.
\end{enumerate}

\textbf{AB:}
The most common example of AB data was human-readable
descriptions of service roles or behaviours. The only exception was
application P07, which included structured OpenAPI documentation with
descriptions embedded as metadata—something our extractors could parse.
However, even in this case, the descriptions still had to be written manually by developers,
they were merely stored in a location that our study methodology allowed extractors
to access.
We do not consider our inability to generate such high-level
descriptions to be a shortcoming of ModARO; such information
is not typically considered within the purview of static analysis,
and no existing reconstruction research attempts
to automatically collect this type of information.
However, recent advances in large language models (LLMs)
present new opportunities; there is no technical limitation preventing
an extractor from using an LLM to generate descriptive summaries
from source code, documentation, or system artefacts.
Such integrations could complement our
extractor design and are a promising direction for future work.
Other information we have categorised as AB are high-level categorisations such
as the design language of a user interface, which may not be information visible
in the code.

\textbf{IR:}
The only documented data that we lacked the time or resources to
extract were the URL parameters accepted by
the HTTP endpoints and the response body schemas from those endpoints for P10.
The reason we deemed these too resource intensive was that there was no central location  
that specified these details; the information would have needed to be pieced
together and inferred from dynamically created and modified objects in the
Ruby codebase.
We also lacked any prior
knowledge or experience with Ruby, which would have made the development of an
extractor even more time-consuming. However, it is possible that a greater understanding of
Ruby would have revealed a simpler solution.
While this case presents a possible limitation in what our approach can feasibly
extract, the challenge stems from the inherent difficulty
of statically reconstructing a data structure by detecting its use across a dynamically
typed codebase.
As such, this is not indicative of a flaw in our approach but a rather
manifestation of the inherent weaknesses and limits of static analysis.

\textbf{UN:}
Unknown data was the information that we did not extract, not because it was too difficult
or because it was not present in the code, but because we could not determine what to
look for in the codebase. This was either because we were unable to interpret the
meaning of the documentation, as with message broker usage for Lelylan;
or because we lacked the understanding of the source code and/or the frameworks
used to know what to look for, as with the endpoint authentication for piggy-metrics.
These issues would not occur if the developers themselves were configuring the reconstruction
as they would necessarily already know these details.
There is no indication that these two cases would have altered
the findings of the study had we understood what to extract.

\subsubsection{Accuracy of our Output} \label{section:accuracy}
The purpose of this study is to test whether our approach can extract
the data documented by developers of the open source microservice projects.
Evaluating the accuracy of the extracted data is not within the
scope of this research. The reason for this is twofold:
\begin{enumerate}
    \item The specific accuracy of the output is largely a product
        of the code of the individual extractors. If an extractor
        has, for example, a buggy regex expression that misses some edge cases,
        then that is a problem with the extractor, not our overall
        approach.
    \item Evaluating the accuracy of our approach requires a source of truth
        to compare with our output. Unfortunately, the documentation
        in these projects is not reliable; we encountered numerous
        cases where the documentation appeared to be incorrect or out-of-date.
        We lack the resources to properly validate and correct the documentation
        for all these projects.
\end{enumerate}
However, while we did not systematically test the accuracy of the output,
the act of writing extractors required comparing samples of the output
against the documentation, so we did detect some discrepancies
during the study.
When we encountered these discrepancies, we attempted to identify the
problem and fix any flaws in our extractor logic, writing new ones if necessary.
However, there were some cases where discrepancies could not be resolved;
we have categorised the reasons we could not do so and marked these cases
in Table~\ref{table:documentation-data} with the following symbols:

\begin{enumerate}
    \item [\textbf{*}] The documentation was incorrect.
    \item [\textbf{$\neq$}] Data was encoded differently in code and documentation (e.g. differently formatted names).
    \item [\textbf{?}] We lacked the understanding of the codebase to determine if the problem
                was the documentation or our extractors.
\end{enumerate}
Anything that did not fit into these categories would be
recorded under one of the categories in Section~\ref{section:doc-not-reconstructed}.
    
Crucially, none of these issues indicate a flaw in the design of our approach.
They are the result of human error; ours, when writing extractors,
or the original developers, when documenting the system.
It is possible that a mistake on our part, or a lack of
understanding of the codebase, concealed an edge-case scenario
that challenges our approach. However, we do not see any indication
that this is likely.
In addition, the ``*'' cases provide examples of the struggle
to keep documentation correct and up-to-date. P4, P5, and P8 are all
demonstration projects designed for education and testing; they
are not subject to the same evolutionary and maintenance pressures of
a real production project, and yet they have seemingly failed to keep their
documentation updated.

\subsubsection{Programming Languages} \label{section:programming-languages}

\begin{table}
    \centering
    \caption{Programming languages used by the applications.}
    \label{tab:application-languages}
    \includegraphics[width=\linewidth]{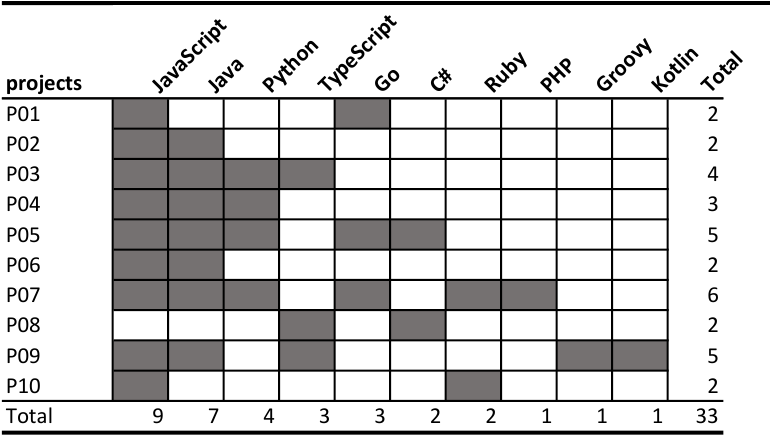}

\end{table}

\begin{table}
    \centering
    \caption{
        Programming languages used by 373 microservice projects as reported in~\cite{AmorosodAragona2024}.
        Templating and markup languages have been removed. Note that projects can use more than
        one language, so the sum of each column is more than the totals.
    }
    \label{tab:dataset-languages}
    \resizebox{.6\linewidth}{!}{
        \begin{filecontents*}{data/dataset-languages.csv}
Language,Projects,% of Total
javascript,120,32.2%
python,90,24.1%
java,79,21.2%
shell,55,14.7%
typescript,53,14.2%
go,37,9.9%
c\#,28,7.5%
php,27,7.2%
ruby,15,4.0%
scala,7,1.9%
elixir,7,1.9%
rust,6,1.6%
kotlin,6,1.6%
c++,4,1.1%
perl,3,0.8%
c,3,0.8%
common lisp,2,0.5%
f\#,1,0.3%
standard ml,1,0.3%
batchfile,1,0.3%
powershell,1,0.3%
Total,373,100.0%
\end{filecontents*}
\csvreader[
            tabular=l r r,
            table head=   \toprule Language & Projects & \% of Total \\ \midrule,
            late after last line=\\\bottomrule,
            before line=\ifcsvstrcmp{\csvcoli}{Total}{\\\midrule}{},
        ]{data/dataset-languages.csv}{}{\csvcoli & \csvcolii & \csvcoliii}
    }
\end{table}

To facilitate architecture reconstruction across a range of microservice projects,
ModARO must be able to extract information from many different
programming languages. Given that our extractor approach allows for arbitrary
execution of code, there are theoretically no limitations in the languages that our
approach can support. However, it is still worth determining if this assumption
holds up in practice, especially considering that during the study we were
restricting ourselves to only using RegEx for code parsing.

We detected the programming languages used by the projects with the
\texttt{cloc}\footnote{\url{https://github.com/AlDanial/cloc}} tool.
Table~\ref{tab:application-languages} shows the languages used by the
study projects. Markup, templating, data format, and build/shell languages
(HTML, CSS, JSON, Bash, Gradle, etc.) have been excluded.
Note that projects can and do use more than one
language, which demonstrates why a polyglot approach is necessary.
JavaScript was the most common language, which can be attributed to
the prevalence of web UIs,
with Java and Python being the next two most common.

We were able to reconstruct information about the applications
for each of the languages encountered, despite some of the languages (e.g. Java and Python)
having notable differences in syntax, typing, and compilation (or lack thereof).
The use of a shared model to store the extracted data, which all extractors agreed
to, also meant that there were no issues when collecting the same
information from multiple languages at the same time.
For example, the \emph{trainticket-spring-requests}\footnote{\url{https://github.com/oscar-manglaras/ModARO-Github-Study-Extractors/blob/master/extractors/dist/application/06/spring-requests.js}}
and \emph{trainticket-python-requests}\footnote{\url{https://github.com/oscar-manglaras/ModARO-Github-Study-Extractors/blob/master/extractors/dist/application/06/python-requests.js}}
extractors both detected
HTTP requests made within the \emph{train-ticket} project. As both
used the same schema to represent HTTP requests in the model, they were both able
to be run at the same time and added request data to the model
for Java and Python microservices, respectively.
The lack of notable challenges for these languages is a promising
sign that our approach is supportive of polyglot microservice systems.

To determine how representative the programming languages of our projects are,
we compared them against
a recently published dataset~\cite{AmorosodAragona2024} of
378 publicly available microservice projects.
This dataset is over six times larger than
the dataset we used to select our study projects, and had it been available at the time,
we would likely have used it instead.
After removing templating and markup languages,
which filtered out 5 projects, we attained the results in Table~\ref{tab:dataset-languages},
from which we can conclude that the languages used in our study are broadly representative
of language usage within the larger dataset.

\subsubsection{Technologies}
\label{section:rq1/technologies}

\begin{table}
    \centering
    \caption{
        Technologies from which extractors collected information during the study.
        This includes both information obtained from files tied
        to particular technologies
        (\texttt{docker-compose.yml} files for Docker Compose, Java source files for Java,
        \texttt{package.json} files for NodeJS, etc.), and analysis of how specific
        libraries and frameworks (Java Spring, ExpressJS, Tornado, etc.) are used within
        those files.
    }
    \includegraphics[width=\linewidth]{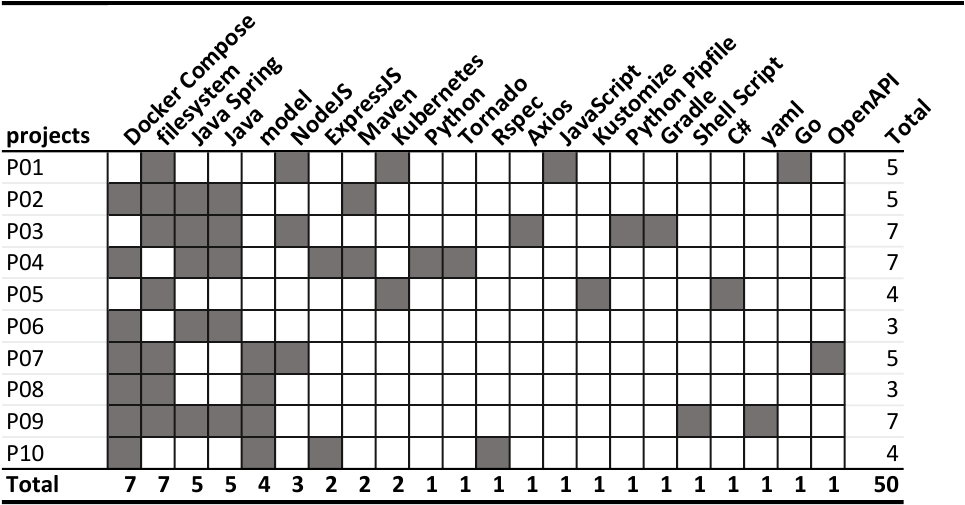}
    \label{fig:technologies}
\end{table}
Table~\ref{fig:technologies} shows the various technologies
(programming languages, packaging formats, deployment configurations, etc.)
from which the information was collected during the study.
To generate this data, we first identified the technology or technologies from which each extractor
sourced its information.
Then, for each project, we gathered all its
extractors to determine which technologies they used.
We are only counting technologies from which information was sourced, 
not technologies about which we collected information; for example,
an extractor that detects the usage of HTML by looking at
file extensions would be classified under `filesystem', but not `HTML'.
The `filesystem' technology refers to extractors that solely collect information
from directory structures or file extensions, the `model' technology refers to
extractors that solely use information already in the model entity received as input,
and the other technologies can be found with a web search.

We can see from this table that the information was sourced from a range of
technologies, including different deployment tools (Docker Compose and Kubernetes),
web frameworks (Java Sping and ExpressJS), and dependency/build tools (Maven and Gradle).
Thus, there is no indication that our study was skewed by reliance on any single technology.

\begin{tcolorbox}[rq]
\textbf{RQ1 Answer:}
Our approach was successful at extracting the documented data from the ten
projects in the study, with no issues arising from the range of languages
and technologies in use.
We could not extract all documented information,
but these cases were due to the inherent limitations of static analysis or
unclear or inaccurate documentation.
\end{tcolorbox}

\subsection{\textbf{RQ2 - How reusable are extractors across different
microservice projects and what factors limit their reuse?}}
\label{section:rq2}

Our extractor design is intended to support re-use across projects or services that
rely on the same underlying technologies. But how effective is this design in practice?
To what extent were extractors actually reused during the study,
why could some extractors not be reused, and are there any changes to our approach
that would improve their reusability?
To answer these questions, we categorised which extractors were
not reusable and the reasons why. To test the accuracy of our
categorisations, a different author, who has 5 years of experience
designing and developing industry microservice architectures, independently categorised
11 extractors (15\%).
We would have reviewed more, but they were unable to due to time constraints.
We found one miscategorisation (Cohen's Kappa 0.79); an extractor
miscategorised as project specific because the original researcher
failed to recognise it was using a common library.
The findings in this paper reflect the corrected categorisation.

\subsubsection{Reuse in Practice}
We developed 71 extractors over the course of the study to analyse 10 GitHub projects.
Of these, 44 (62\%) were targeted at specific technologies and are hence theoretically reuseable,
while the remaining 27 (3815\%) ended up being project-specific; tightly coupled to individual projects.
These \emph{project-specific} extractors needed to collect data
using unique code conventions, file structures, or naming schemes.

Of the reusable extractors, only 34\% ($\frac{15}{44}$) were actually
used multiple times in the study to collect documented information.
In addition, 61\% ($\frac{27}{44}$) \emph{could} have been reused
on more than one project to gather additional data beyond what was documented.
However, these reuse rates are primarily a reflection of which technologies appear
in more than one study project; they are not indicative
of how genuinely reusable the extractors are.
In addition, many extractors could plausibly be split into
smaller extractors or combined into larger ones.
For example, when detecting if specific programming languages were
being used, 
we chose to detect each with a separate extractor; a single, larger, extractor
for all languages would have been just as viable.
For these reasons, the number of reused extractors is not a particularly useful
metric.

Within the set of reusable extractors, eight had to
be modified to support edge cases encountered in new projects.
These changes were backward-compatible, preserving the functionality of extractors
for earlier use cases, and thus we still consider them reusable.
However, a separate subset of extractors required changes that were not backward-compatible.
These extractors still need to extract the same type of information
from the same technologies, but differences
in project configuration or implementation details prevented direct reuse.
We refer to these as \emph{variant extractors} and will discuss them separately.

\subsubsection{Project-Specific Extractors} \label{section:project-extractors}

\begin{table}
    \centering
    \caption{Reasons why an extractor was made project-specific.}
    \label{tab:project-extractor-categories}
        \begin{filecontents*}{data/project-extractors-counts.csv}
Category,Count
Collection of project-specific information,14
Use of code conventions to simplify logic,7
Custom directory structure,5
Non-conformant file format,1
\end{filecontents*}
\csvreader[
            tabular=l r,
            table head=   \toprule category & count \\ \midrule,
            late after last line=\\\bottomrule,
        ]{data/project-extractors-counts.csv}{}%
        {\csvcoli & \csvcolii}
\end{table}

Of the 71 extractors developed during the study, 28 (39\%) were inherently
project-specific. To better understand this limitation,
we categorised each of these extractors based on why they could
not be reused.
The resulting categories and their frequencies are shown in
Table~\ref{tab:project-extractor-categories}.
The extractor categorised under \textit{non-conformant file format}
was processing a Docker Compose file that used invalid syntax.
The other categories represent more common scenarios.

\textit{Collection of project-specific information:}
The majority of these extractors were developed to collect information that was unique to a
specific project or that was stored in unique ways. This included,
for example, custom categorisations defined by developers
(such as tagging services as UI frontends or proxy servers) or extracting details like port
numbers from arbitrary configuration files or custom build scripts, and in some cases
this meant modifying an existing extractor to gather additional project-specific
information from the code it was already analysing.
In such cases, reuse is not realistically feasible and we argue that this
does not reflect a shortcoming of our extractor design.
On the contrary, the presence of these extractors highlights a core strength of our approach:
enabling the creation and integration of custom extractors when necessary. ModARO supports
the ad hoc development of extractors to meet specialised data collection needs,
something we consider a design success rather than a failure.

\textit{Use of code conventions to simplify logic:}
In some projects, developers adhered to naming or structural conventions
that made it easier to extract otherwise complex information from the code.
All of these extractors were designed to identify communication relationships
between services by identifying classes and function calls that represented
proxy routes, HTTP requests, or asynchronous messaging.
These abstractions were either developed in-house or provided
by third-party frameworks like Java Spring.
However, even when using common frameworks, these extractors were still project-specific
due to the need to process string-based identifiers such as HTTP request paths
or RabbitMQ topic names. The primary challenge lies in the variety of ways these
strings are constructed. Some projects used static string literals, while others
built identifiers dynamically using function calls,
concatenation, or a mix of constants and variables.
Handling all of these possibilities
in a general-purpose extractor would require complex, language-specific static
analysis, and even that might not capture values derived
from runtime behaviour or environment variables.
These extractors highlight the potential value of consistent code conventions
for static analysis;
by adhering to predictable patterns, developers can significantly simplify
what would otherwise be a complex or infeasible extraction task.
Unless projects adopt the same conventions, we do not see how these
extractors could have been made re-usable.

\textit{Custom directory structure} \label{section:custom-filepaths}
A notable cause of project-specific extractors was the use of non-standard filepaths
or directory layouts. In these cases, reusable extractors failed simply because expected
files were located in unexpected places or named unconventionally,
e.g. custom named docker-compose files.
These discrepancies are not mistakes by developers;
many tools, including Docker Compose,
explicitly support arbitrary filenames and rely on users to specify the
correct paths at runtime when so. Consequently, extractors that assume standard conventions
are inherently limited: they cannot handle valid configurations that deviate from the defaults.
This category exposes a key limitation in our current extractor design. Simply importing
the appropriate extractor is not sufficient; to fully support flexible technologies,
extractors must be able to either detect variations in file structures (which may not
be viable in practice) or accept external
configuration (e.g., via parameters or environment variables). Without this capability,
extractors remain brittle in the face of perfectly valid but non-standard project setups.

\subsubsection{Variant Extractors}

Most variant extractors were a subset of the project-specific extractors discussed
in the previous section. They were created by duplicating existing extractors and modifying
them to handle custom filepaths or to gather additional project-specific
information from the same code already being analysed.

However, two notable exceptions stood out: the \emph{docker-compose-networking} extractor,
which gathered network port information, and the
\emph{docker-compose-environment} extractor,
which captured environment variables. Both were still reusable in principle but were
separated from the base \emph{docker-compose} extractor, not due to parsing limitations,
but because the values they collected occasionally conflicted
with what developers wanted documented.
Since ModARO does not provide control over what data an extractor
should include or ignore, the only workaround was to split the original extractor
into three variants, each duplicating roughly 80–90\% of the code.
It is reasonable to expect that extractors for other technologies may also require
selective data extraction.
The ability to tell extractors what data to output
would improve reusability and help avoid unnecessary duplication.
This reinforces the suggestion made in the previous subsection to allow
passing configuration values to extractors.

\begin{tcolorbox}[rq]
\textbf{RQ2 Answer:}
Of 71 extractors, 44 were reusable while 27 were not due to collecting project-specific
information, relying on code conventions, or accounting for custom directory structures.
Extractors would require fewer project-specific modifications if they
could be passed configuration options at runtime.
\end{tcolorbox}

\subsection{\textbf{RQ3 - How well does our approach support distributed architecture reconstruction across multiple microservice repositories?}}
\label{section:rq/distributed-reconstruction}

\begin{figure*}[!t]
    \centering
\begin{minted}{typescript}
import { SuperExpressive, arrayAppend, createLink, doAnyFilesContain, registerExtractor, schemas, globEscape } from "api";

const eurekaServerRegex = SuperExpressive()
    .char('@')
    .zeroOrMore.whitespaceChar
    .string('EnableEurekaServer')
    .wordBoundary
    .toRegex();

const eurekaClientRegex = SuperExpressive()
    .char('@')
    .zeroOrMore.whitespaceChar
    .string('EnableEurekaClient')
    .wordBoundary
    .toRegex();

async function extractor(microservice) {
    const javaFilesGlob = `${globEscape(microservice.$path)}/**/*.java`;
    const isServer = await doAnyFilesContain(javaFilesGlob, eurekaServerRegex);
    const isClient = await doAnyFilesContain(javaFilesGlob, eurekaClientRegex);
    
    if (isServer) {
        arrayAppend(microservice, 'tags', 'EurekaServer');
    }

    if (isClient) {
        arrayAppend(microservice, 'tags', 'EurekaClient');

        // creates a retroactive link using a json-schema that matches
        // a microservice entity with a EurekaServer tag
        arrayAppend(microservice, 'dependencies', await createLink('/microservices', {
            type: 'object',
            properties: {
                $TYPE: { const: 'microservice' },
                tags: schemas.arrayContaining({ const: 'EurekaServer' }),
            },
            required: ['$TYPE', 'tags']
        }));
    }

    return microservice;
}

registerExtractor(extractor, schemas.presets.localMicroservice)
\end{minted}
    \caption{
        An extractor (based on \textit{spring-eureka} from the study\textsuperscript{*})
        that uses retroactive links
        to create a dependency between Eureka clients and servers.
    }
    \footnotesize\textsuperscript{*} \url{https://github.com/oscar-manglaras/ModARO-Github-Study-Extractors/blob/master/extractors/dist/technology/spring/eureka.js}
    \label{fig:link-example}
\end{figure*}

\begin{figure*}[t]
    \centering

    \begin{subfigure}{0.3\textwidth}
        \includegraphics[width=\linewidth]{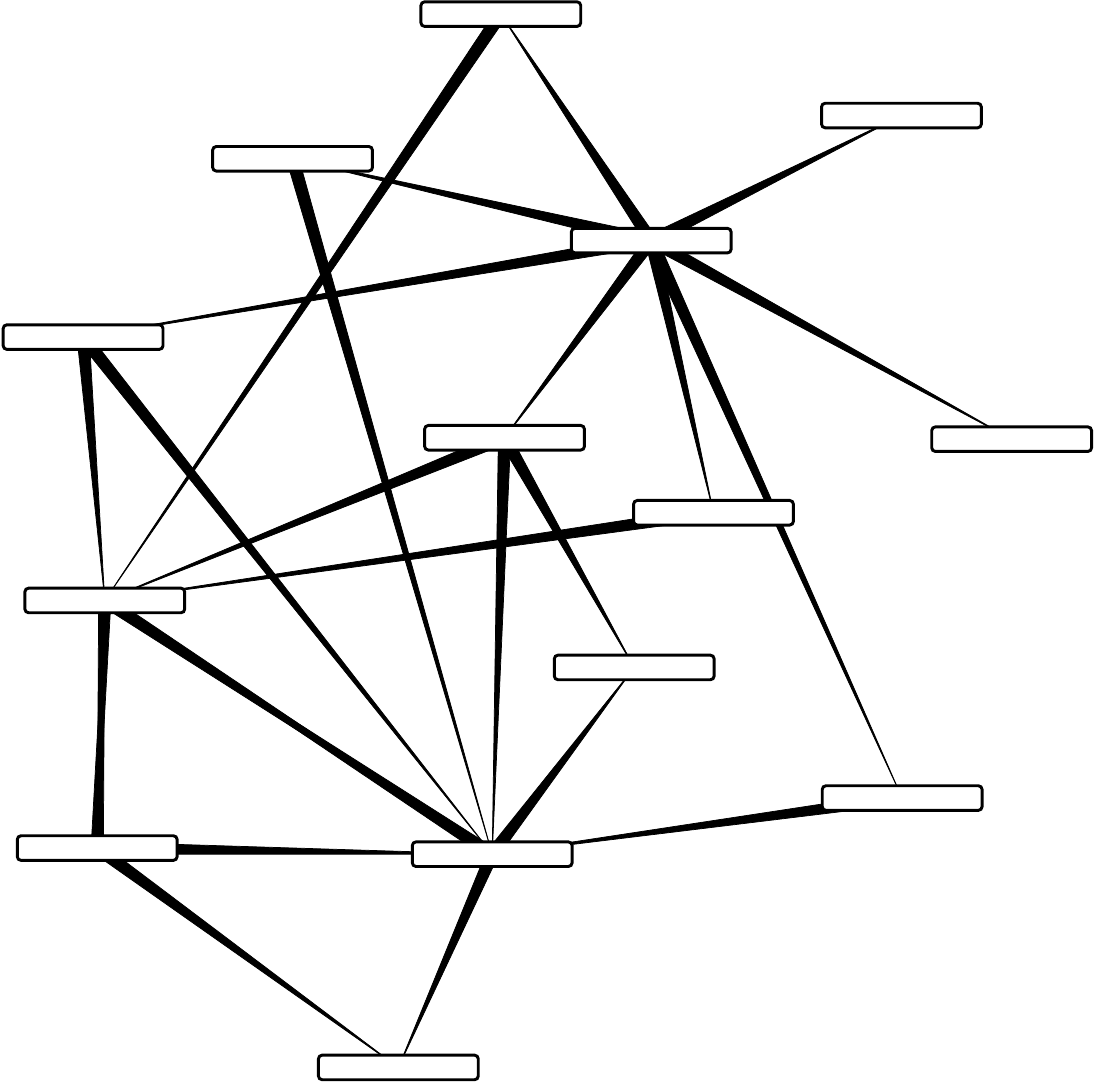}
        \caption{P1 - StaffJoy}
        \label{fig:top-staffjoy}
        \label{fig:staffjoy}
    \end{subfigure}
    \hfill
    \begin{subfigure}{0.3\textwidth}
        \includegraphics[width=\linewidth]{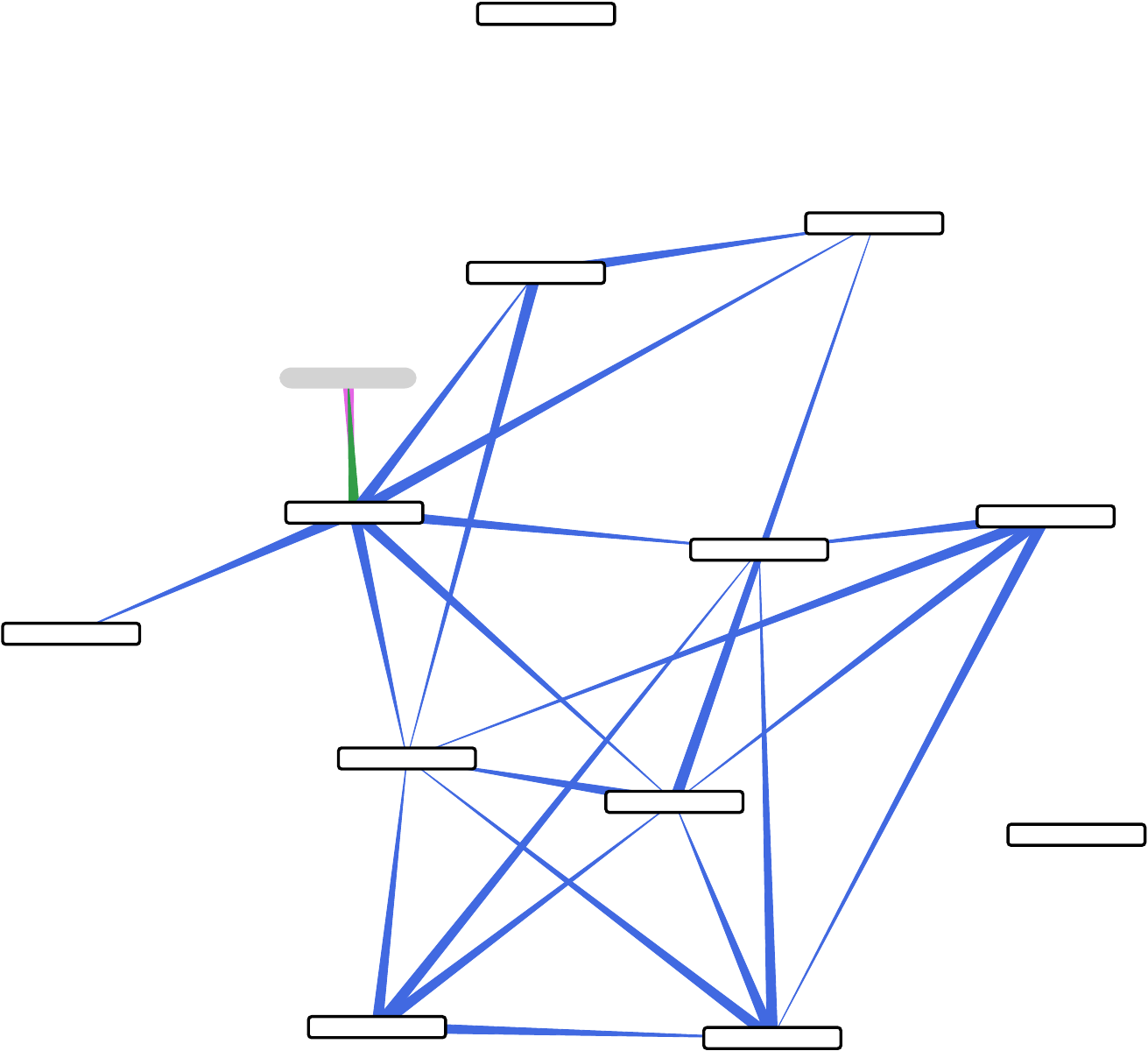}
        \caption{P2 - Cloud Native Strangler Example}
        \label{fig:second}
        \label{fig:cloudnativestrangler}
    \end{subfigure}
    \hfill
    \begin{subfigure}{0.3\textwidth}
        \includegraphics[width=\linewidth]{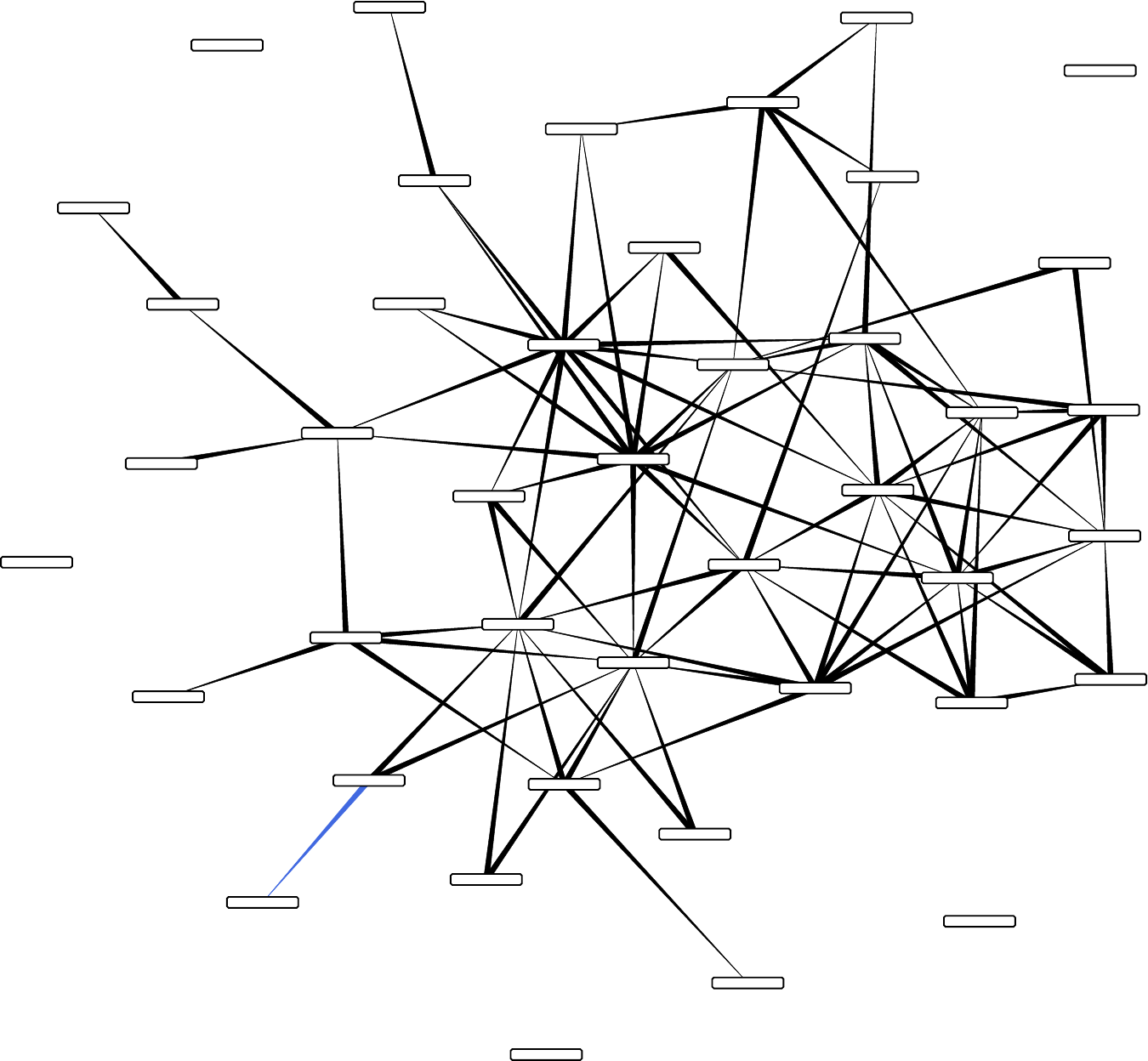}
        \caption{P4 - Train Ticket}
        \label{fig:third}
        \label{fig:trainticket}
    \end{subfigure}
    \vspace{0.5cm} %

    \begin{subfigure}{0.3\textwidth}
        \includegraphics[width=\linewidth]{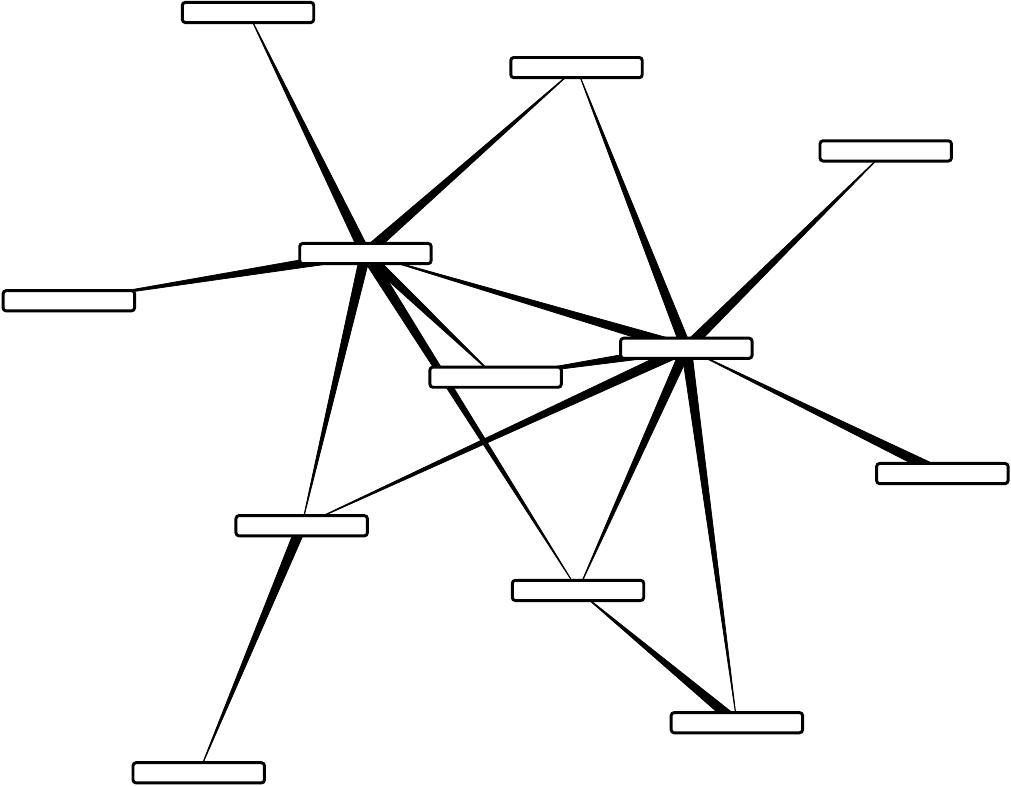}
        \caption{P5 - Hipster Shop}
        \label{fig:fourth}
        \label{fig:hipstershop}
    \end{subfigure}
    \hfill
    \begin{subfigure}{0.3\textwidth}
        \includegraphics[width=\linewidth]{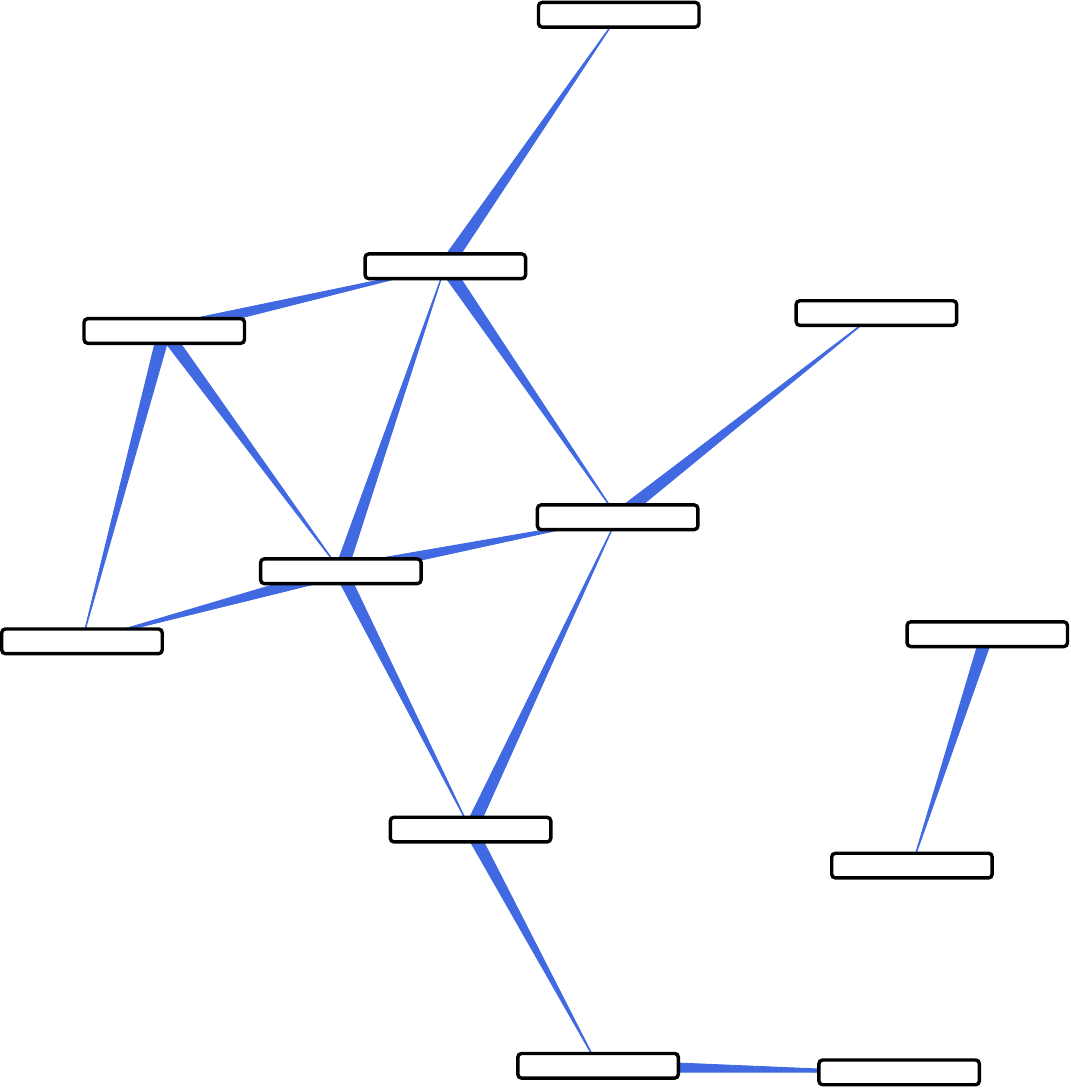}
        \caption{P7 - Photo Uploader}
        \label{fig:fifth}
        \label{fig:photouploader}
    \end{subfigure}
    \hfill
    \begin{subfigure}{0.3\textwidth}
        \includegraphics[width=\linewidth]{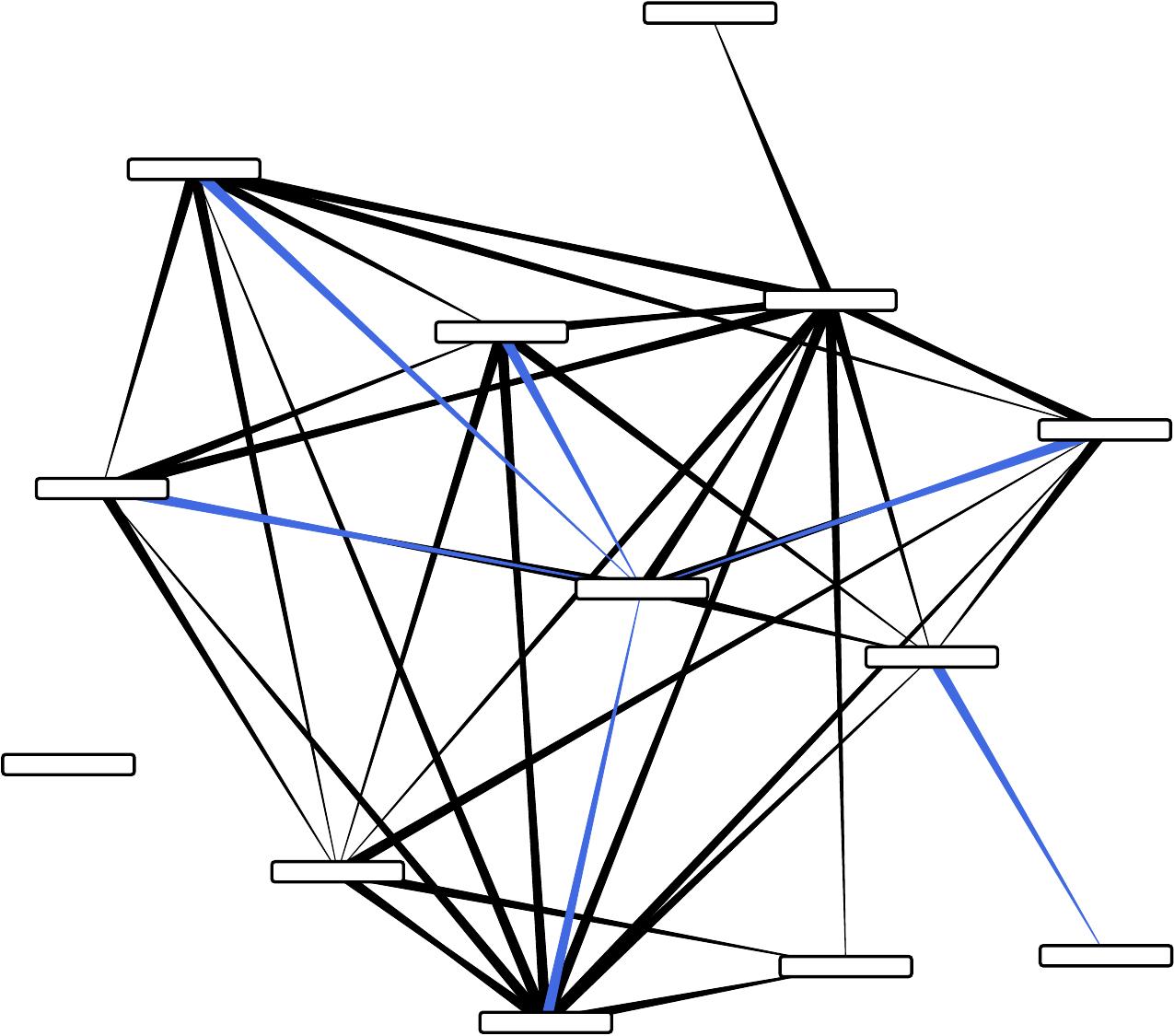}
        \caption{P9 - Spinnaker}
        \label{fig:sixth}
        \label{fig:spinnaker}
    \end{subfigure}
    
    \caption{
        Inter-service connections found in the study projects.
        Black links are synchronous requests (e.g. HTTP), green and purple are publications
        and subscriptions (respectively) to an asynchronous channel (the grey nodes),
        and blue are for any other type of dependency (e.g. from \texttt{depends\_on} in docker compose files).
        Labels have been removed for visibility purposes.
    }
    \label{fig:topologies}
\end{figure*}

To answer this question, we must look at two issues:
is our approach able to collect information distributed across multiple
repositories, and does our approach allow extractors to be reused
in a distributed context?

\subsubsection{Can extractors be reused for multi-repo projects  }
During the study, we tracked which extractors were created, reused, or
modified for each project. We also recorded when a reusable extractor
needed to be modified to work with a new project in a way that made it
incompatible with previous projects (referred to as \emph{variant extractors}
in Section~\ref{section:rq2}). The two multi-repo study projects required
7 reusable extractors to collect the documented information, of which 3
were created for earlier mono-repo projects.
None of these extractors required any modifications or custom behaviour to support
the multi-repo projects; the only reusable extractor that needed to be modified was
(\textit{spinnaker-endpoints}\footnote{\url{https://github.com/oscar-manglaras/ModARO-Github-Study-Extractors/blob/master/extractors/dist/application/11/endpoints.js}}),
a variant extractor that was changed to support
Groovy and Kotlin code syntax instead of Java, which has nothing to do with distributed codebases.
As such, there is no indication that distributed
architecture reconstruction impaired extractor reuse.

We attribute the successful reuse of extractors in mono- and
multi-repository contexts to the structure of our model and
the way our reconstruction algorithm sends
model entities to extractors. An extractor targeting a microservice
is only given the model entity for that service, which, even for
mono-repo projects, encourages extractors
to collect information from only the source code of the target service.
Figure~\ref{fig:link-example} shows an example of an extractor that
creates links between services using retroactive linking
without having to access the code of other services.

\subsubsection{Can we collect/aggregate information}
In Section~\ref{section:rq1}, we reported the information that
extractors could not collect and why.
Of the uncollected information, none were due to the
distributed nature of the project. One multi-project (P09) had inaccuracies
in the output, but we were unable to determine the cause due to
the complexity of the codebase and the unreliability of the documentation.
The retroactive links were also a success;
figure~\ref{fig:topologies} shows the connections between services in 
study projects for those that documented them (P3 also documented connections,
but was excluded because it only had two links).
All synchronous and publish/subscribe connections (non-blue links)
were specified using retroactive links by a microservice-specific
extractor, with the exception of P5 (Figure~\ref{fig:hipstershop}),
which extracted the synchronous connections using architecture-level deployment
configuration files. 
In contrast, dependency links (blue) were detected using
deployment-level information, with the exception of P9 (Figure~\ref{fig:spinnaker}).
Detecting and specifying links at the deployment level still works in a multi-repo
environment, one just needs to run the reconstruction approach on the
deployment configuration and include those results in the aggregation.

However, while conducting the study, we identified a limitation of our
link implementation; when the link schema itself relies on information
that is present in a remote repository. An example is an HTTP request in
which the domain of the target microservice is stored in an environment
variable set in a different code repository.
Project P10 did record domains in this way, but requests to them were
undocumented, which is why this issue does not appear in the data for RQ1.
To avoid hard-coding values into the extractors of multiple repositories
(which our linking approach is intended to avoid), links must be made
more powerful and expressive; allowing the schema values that define links
to also be resolved retroactively. 
This shows that the links need more functionality than we initially assumed
(and that JSON schema alone is insufficient),
but does not fundamentally change the nature of our approach or the viability of
distributed architecture reconstruction.

\begin{tcolorbox}[rq]
\textbf{RQ3 Answer:}
Extractors were reused successfully across mono- and multi-repo projects,
and retroactive links were resolved without issues.
However, to define accurate links, some projects require link schema values to themselves be
resolved retroactively.
This scenario would require
extending link functionality beyond our initial assumptions,
but does not fundamentally change our approach.

\end{tcolorbox}

\section{Software Developer User Study} \label{section:rq/user-study}
\label{section:user-study}
\begin{figure*}[t]
    \centering
    \begin{subfigure}{\textwidth}
\begin{minted}{json}

{
  "identifier": "ClassOrInterface",
  "pattern": ".*",
  "auxiliary_pattern": "com.piggymetrics.#{name}",
  "essential": true,
  "subpatterns": [
    {
      "identifier": "Annotation",
      "pattern": "@EnableDiscoveryClient",
      "subpatterns": [],
      "essential": true
    }
  ],
  "callback": "let microservices = ctx.get_or_save(\"microservices\", []); let name = ctx.get_variable(\"name\")?; let i = 0; while i < microservices.len() { if microservices[i].name == name { microservices[i].discoveryClient = true; } i += 1; }"
}
\end{minted}
        \caption{
            Normal analyser. We assume that another analyser
            has already created the array of microservices
            with the correct names.
        }
    \end{subfigure}
    \begin{subfigure}{\textwidth}
        \centering
\begin{minted}{json}
{
  "$MODEL_TYPE": "microservice",
  "identifier": "ClassOrInterface",
  "pattern": ".*",
  "auxiliary_pattern": ".*",
  "essential": true,
  "subpatterns": [
    {
      "identifier": "Annotation",
      "pattern": "@EnableDiscoveryClient",
      "subpatterns": [],
      "essential": true
    }
  ],
  "callback": "ctx.save(\"discoveryClient\", true);"
}
\end{minted}
        \caption{
            Hybrid analyser. We assume that a separate
            extractor is creating the array of microservices
            and setting the subpaths of the microservice
            repositories.
            Note that there is a \texttt{\$MODEL\_TYPE} field in the analyser,
            this corresponds to the \texttt{\$TYPE} field of model entities,
            and allows specifying if the analyser should run on individual microservices
            (``microservice'') or the full repository (``\$MODEL'').
        }
    \end{subfigure}
    \caption{
        An example of a trivial ReSSA analyser that
        detects \texttt{@DiscoveryClient} annotations
        for extractors, and its hybrid equivalent.
    }
    \label{fig:hybrid-example}
\end{figure*}

\subsection{Study Overview}

In addition to validating our design assumptions against microservice projects,
we also wanted to evaluate the more subjective aspects of ModARO, such
as usefulness and ease-of-use. To do so, we conducted a study with industry practitioners:
we gave them our implementation and asked them to complete a small number
of data extraction tasks. We then collected structured feedback.

To provide a point of comparison, we had participants complete data extraction
tasks using the ReSSA
tool\footnote{\url{https://github.com/cloudhubs/source-code-parser/tree/main}}~\cite{Schiewe2022a}.
We selected this tool because it was the only other static reconstruction approach
that we were aware of at the time
that supports arbitrary data extraction in a language-agnostic fashion.
ReSSA supports defining analysers that derive code structures from a
language-agnostic abstract syntax tree (LAAST). This involves first generating
language-specific concrete syntax trees (CSTs) and then transforming them into a corresponding LAAST.
Supporting new languages requires the use or creation of a parser
that can create a CST for the language, and code to translate
that CST into the equivalent LAAST. The provided implementation includes
parsers and transformers for Java and C++. To avoid the need to write a new
transformer for a whole language, we only asked participants to use ReSSA on Java code.
Note that this study does not attempt to evaluate the usefulness or effectiveness of ReSSA
and should not be treated as such.

There are key differences between our approach and ReSSA that would
affect the comparison.
Firstly, ReSSA uses full syntax parsing for configured languages,
which allows matches against hierarchical and nested code structures.
In contrast, ModARO only provides lower-level string matching (regex),
with the assumption that more complex parsing
would be provided via additional or external APIs.
Those additional APIs have not currently been created or implemented,
and so tasks that require
extracting data from code may be rated poorly with our approach;
not due to design flaws,
but simply because the intended APIs do not yet exist.
To provide more context around this issue, we created an extractor capable of invoking ReSSA analysers from
within ModARO, allowing for a hybrid method that incorporates ReSSA's analysis
into our approach (see Section~\ref{section:hybrid-approach}).
This third point of comparison allows us to partially disentangle feedback
related to ReSSA's parsing strengths
from feedback about the overall design and UX of each tool.

The other source of potential bias lies in the range of extractable information.
ReSSA is limited to code-based extraction, whereas ModARO
is designed to extract data from arbitrary files.
We know that, in practice, architectural information can reside outside the code,
as demonstrated in Section~\ref{section:rq1/technologies};
if participants were only assigned tasks solvable by both approaches,
they would miss out on one of our key capabilities, potentially undervaluing
the usefulness of our approach.
To mitigate this, we gave participants additional tasks with ModARO
involving non-code files. While this introduces a bias of its own, spending more time
with our approach, we deemed it minor enough to be a worthwhile trade-off.

\subsection{Recruitment and Methodology}
\label{section:methodology/user-study}

The participants were sourced from Swordfish Computing, a software development company
that sponsors this research.
Participation was completely voluntary, 
and participants received their standard pay for the time
they spent taking part in the study. The responses were also kept anonymous
outside the research team and no senior management participated or was
present during the study. To further mitigate any potential bias or power imbalance,
we had Swordfish Computing present an official letter to participants pledging
that there would be no consequences to non-participation.

To be eligible, participants had to meet the following criteria:
\begin{enumerate}
    \item \emph{Has experience working on microservice systems.}
    \item \emph{Has experience with JavaScript,
        the language used for our extractor implementation.}
        Required knowledge includes: functions, conditionals, for loops,
        object creation/referencing, string manipulation.
    \item \emph{Experience with regular expressions.}
\end{enumerate}
Notably, we did not require experience using \emph{Rune}, the Rust-based
scripting language used by ReSSA. As Rune is not widely used, there would
not be enough eligible participants for the study if it was required knowledge.
We acknowledge that this may introduce a bias against ReSSA. To try and mitigate
this issue, we provided participants with a help sheet of standard Rune operations,
and we allowed them to use ChatGPT to help detect syntax errors
and interpret error messages.
The recruited participants all had 3-5 years of software engineering experience
and were a mix of junior and senior employees engaged across 5-6 projects.

The study was conducted with in-person sessions
and consisted of the following activities:
\begin{enumerate}
    \item Participants were introduced to the concept of architecture reconstruction
        and static analysis.
    \item Participants were asked to download a pre-configured development
        environment\footnote{\url{https://doi.org/10.25909/31175716}} with
        the tools already set up and containing pre-written examples and templates.
        API autocomplete for ModARO was provided.
    \item For each approach (ModARO, ReSSA, Hybrid), participants were:
        \begin{enumerate}
            \item Given an API reference sheet.
            \item Given a written explanation of the approach and given time to read and
                ask any questions.
            \item Verbally quizzed on the approach to correct any misconceptions.
            \item Given a list of data extraction tasks to perform in order.
        \end{enumerate}
    \item Participants were given a questionnaire to complete with their feedback.
\end{enumerate}
The tasks were provided on paper and digitally\footnote{\url{https://doi.org/10.25909/31175617}}.
We also included links to a small
number of online resources, such as \url{regex101.com}, to assist with
the tasks. Participants were encouraged to ask any questions to clarify
the instructions or API methods, and to confirm completion of the task.

All steps for a given approach were completed before moving on to the next.
Each session lasted approximately 4 hours, with a break between the extractor and
ReSSA approaches (not included in the 4 hours). Participants were split into two groups, each attending
separate sessions. The sessions were run identically, except for the order in
which the extractor and ReSSA tasks were performed. The hybrid approach
was always completed last.
Due to time constraints, we could not wait until every participant finished
every task. We set a time limit of approximately 60 minutes for
each tool (varying slightly based on how long steps 3a--3c took), after
which participants were asked to stop.

The questionnaire\footnote{\url{}}
contains a mix of Likert scale responses and short-answer
responses. To address our research questions,
we used the \emph{Perceived Usefulness}, \emph{Perceived Ease-of-Use},
and \emph{Intention to Use}
questions from the Technology Acceptance Model 2 (TAM2)~\cite{Venkatesh2000}.
We also included short-answer response questions asking how our extractor approach
could be changed to improve each of these characteristics.
In addition to TAM2, we provided a series of Likert scale questions that asked  
participants to rate how challenging they found different subtasks during the study,
e.g. understanding the concepts, code syntax, data extraction, etc.
Finally, we asked for general feedback about ModARO,
which we analyse with thematic analysis.

\subsection{Hybrid Approach} \label{section:hybrid-approach}
The aim of the hybrid approach is to test how an existing tool with
more advanced parsing capabilities (ReSSA) could be integrated into
ModARO and to test how the advantages our approach provides
to the existing tool would be received by participants.
For this approach, we wrote a custom \emph{ressa-wrapper}
extractor that can be run alongside
other extractors. This extractor can be given simplified ReSSA analysers
using configuration arguments (as suggested in Section~\ref{section:rq2}
and implemented between studies).
The simplified analysers specify the type of model entity to operate on
(the \texttt{\$MODEL\_TYPE} field),
and the wrapper finds model entities of that type with a \texttt{\$path}
field, and runs the analyser separately on each directory (specified by the
\texttt{\$path}). The output of the analyser is then stored directly
in that model entity. Figure~\ref{fig:hybrid-example} shows an example
of a normal ReSSA analyser and its hybrid equivalent.
The primary difference is that the callback string is significantly
simpler in the hybrid version because it does not need to iterate
through the microservice array to find the correct microservice,
the reconstruction algorithm has already passed the correct microservice
entity to the \emph{ressa-wrapper}, which has passed only that
microservice's source code to the analyser.
These hybrid analysers would be usable alongside
normal extractors. Note that we are not attempting to present a new approach,
but to demonstrate how the design of ModARO supports the integration of existing
tools.

\begin{figure*}
    \centering
    \includegraphics[width=\linewidth]{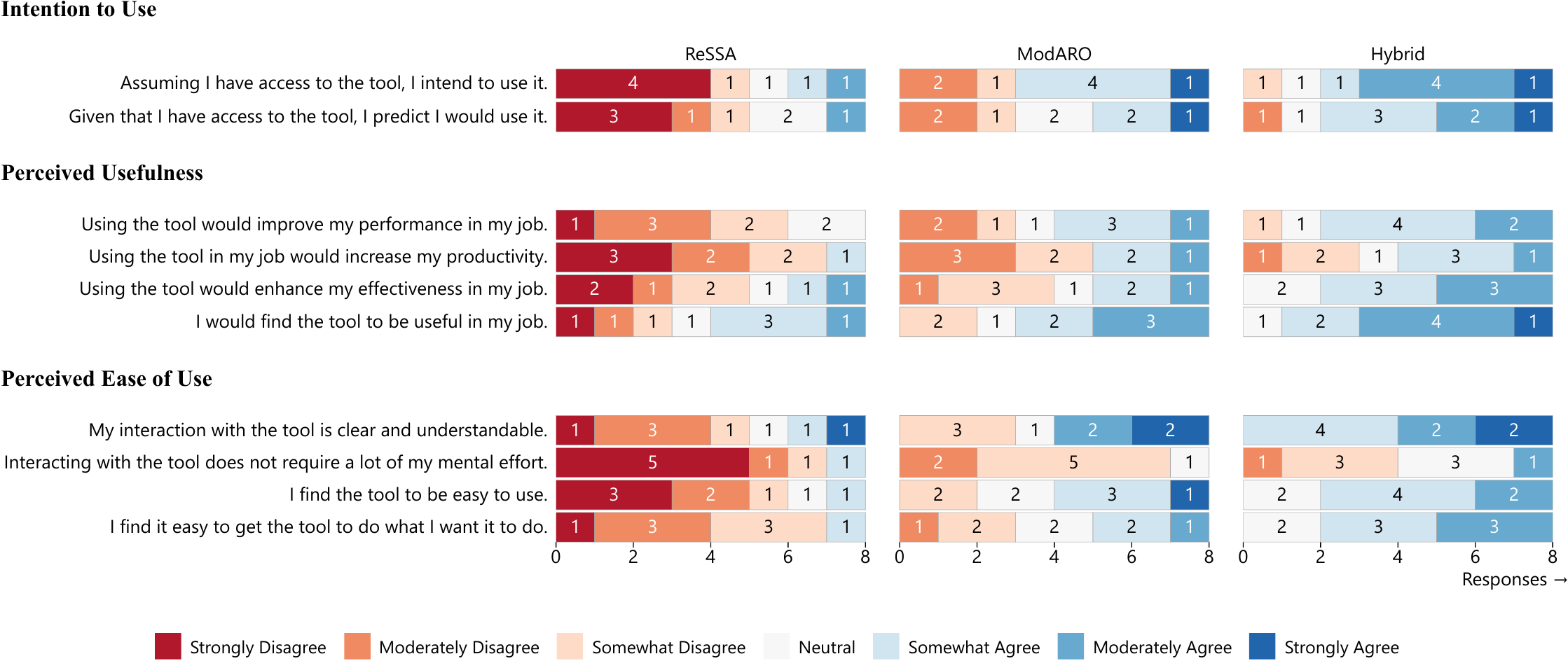}
    \caption{Responses to the TAM2 questions in the user study.}
    \label{fig:usefulness}
    \label{fig:ease-of-use}
\end{figure*}
\subsection{\textbf{RQ4 -- How do practitioners perceive the usefulness
    and usability of our approach, and what factors contribute to or could enhance these aspects?}}

\subsubsection{Intention to Use and Perceived Usefulness}
Figure~\ref{fig:usefulness} shows how participants responded to the TAM2
questions. More than half of the participants answered that they intended to use
ModARO, although there was an even split among the participants who predicted that they
would actually use it. The results for the performance, productivity, and effectiveness
questions were mixed, but only two participants disagreed that ModARO would be useful
in their jobs.
Looking at individual participants in more detail, Figure~\ref{fig:participants-matrix}
shows that every participant rated our approach as better than or equal to the ReSSA
approach for the Intention to Use and Perceived Usefulness questions.
The preference for ModARO was small among most participants,
but they unanimously leaned towards our approach.
In Figure~\ref{fig:usefulness} this is noticeable, with participants responding
\emph{strongly disagree} for questions relating to ReSSA's usefulness, which skews
the graph negatively in comparison to our approach.

Interestingly, the hybrid approach showed notably better results than
both the extractor and ReSSA approaches, despite having no additional
capabilities over either.
Only one participant disagreed that they would use the approach,
and the results were similarly positive for the perceived usefulness questions.
We suspect that the reason was that participants found the hybrid approach easier
or more intuitive to use than the other approaches; but there may also be some
forms of bias.
We will discuss this more when covering the ease-of-use questions.

\subsubsection{Perceived Ease-of-Use}
\begin{figure*}
    \centering
    \includegraphics[width=.95\linewidth]{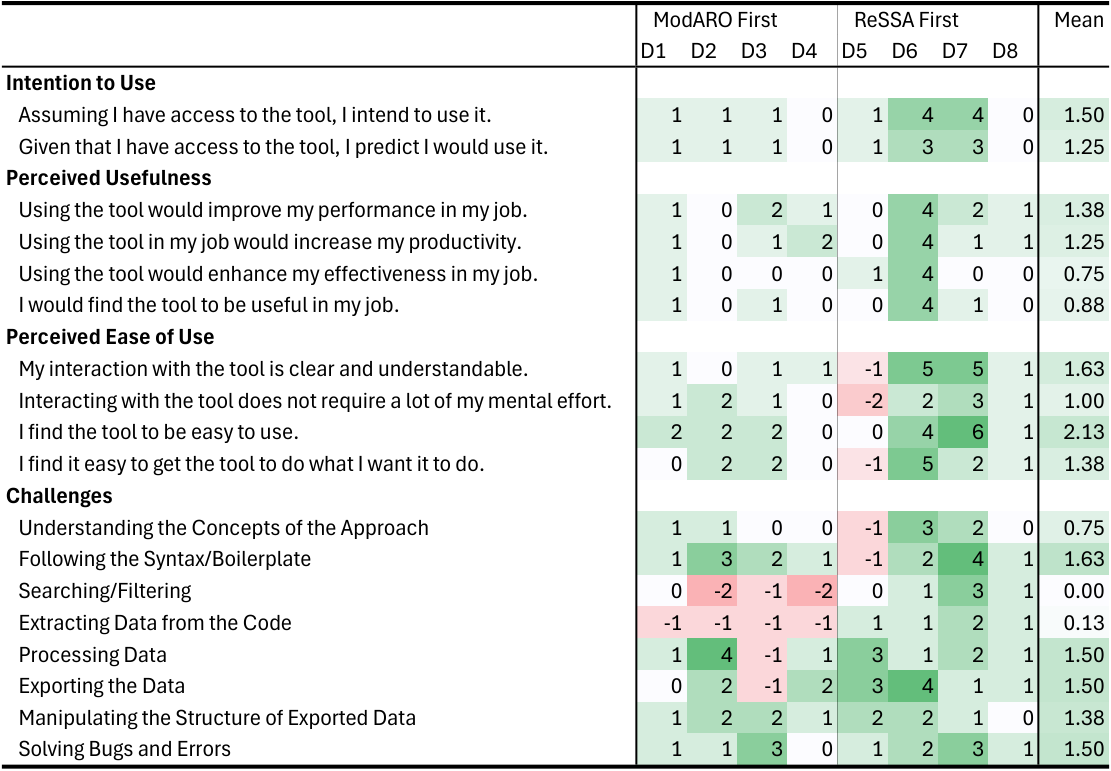}
    \caption{
        Comparison of responses from participants between ModARO and ReSSA.
        The columns represent the participants, the rows the questions
        asked about each tool, and the cells compare their answers.
        Green cells are ones where ModARO was rated better and red where it
        was rated worse.
        The number in each cell is the difference between participants'
        responses for the two tools ($x_{ReSSA} - x_{ModARO}$);
        $x$ is a number from 0 -- n where 0 is the
        worst likert response (\emph{strongly disagree} for TAM2;
        \emph{extremely} for the challenge questions).
        The first four participants (group 1) learnt ModARO first, while the
        latter four (group 2) learnt ReSSA first.
    }
    \label{fig:participants-matrix}
\end{figure*}
\begin{figure*}
    \centering
    \includegraphics[width=\linewidth]{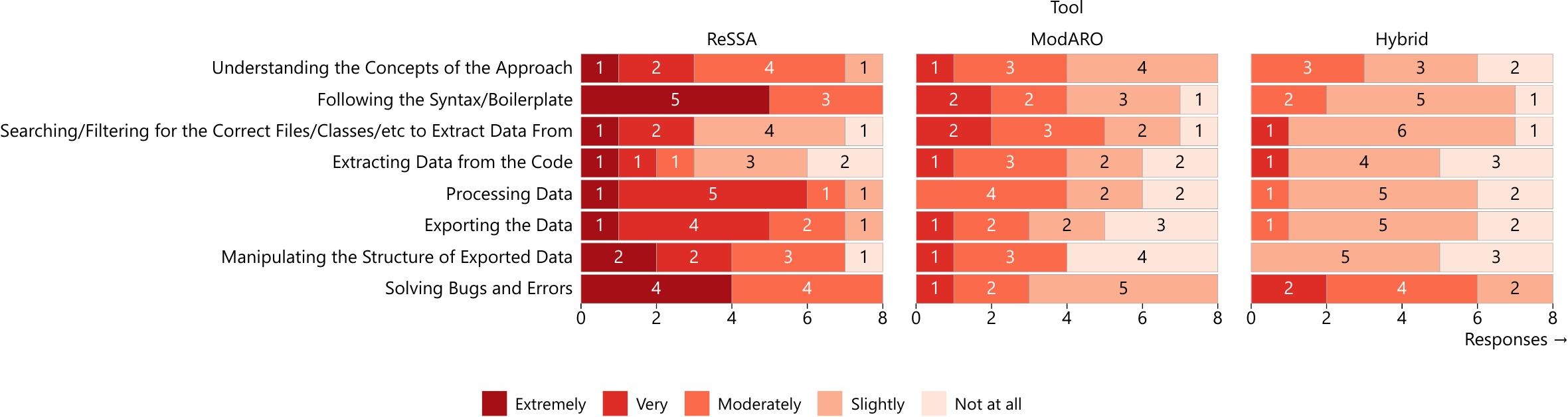}
    \caption{
        How challenging participants found different activities during the
        study. Grouped by tool.
    }
    \label{fig:challenges2}
\end{figure*}

As shown in Figure~\ref{fig:ease-of-use}, more than half of the participants
reported that they found ModARO
easy to use, that they found it easy to get the tool to do what they
wanted, and that their interactions with the tool were clear and understandable.
Almost all of the participants disagreed that the tool
required little mental effort to interact with; however,
this may be due to the complexity of the problem domain and the
mere hour participants had to use each tool, rather than a specific problem
with our approach.
Multiple participants expressed optimism that ModARO would be easier
to use with more time and experience (see Section~\ref{section:theme/learning-curve}),
and they were still able to perform the tasks despite their reported mental effort.
As such, it seems likely that mental effort would decrease with time.

In contrast, the majority of the participants disagreed that the ReSSA approach
was easy to use, yielding notably worse results than our approach.
This can be seen more clearly in Figure~\ref{fig:participants-matrix},
which shows that all but one participant rated ModARO
the same as or better than ReSSA for all the ``ease-of-use'' questions.
We can see that the only participant who found ReSSA easier (D5) is also
the only one who found the concepts and syntax of ReSSA less challenging than
our approach. It is plausible that different people may naturally find
one of the approaches more intuitive than the other, and these responses may
reflect that.

The hybrid approach was the best received, more so than we had expected,
with no participants reporting that it was difficult to use. One possible reason that the hybrid approach
was rated so well is that it required learning fewer new concepts than the other
approaches; participants could mostly re-apply what they already knew about the
extractor and ReSSA approaches. In addition, the tasks that participants performed
with the hybrid approach involved converting ReSSA analysers that they had
already written into the hybrid format; a task notably easier than writing
them from scratch. While we had intended for participants to rate the hybrid
approach based on the full process of writing hybrid analysers, some may have
based their ratings on the porting process only.

\subsubsection{Difficulty breakdown - Likert scale responses}

\begin{figure}
    \centering
    \definecolor{forestgreen}{RGB}{34,139,34}
    \includegraphics[width=\linewidth]{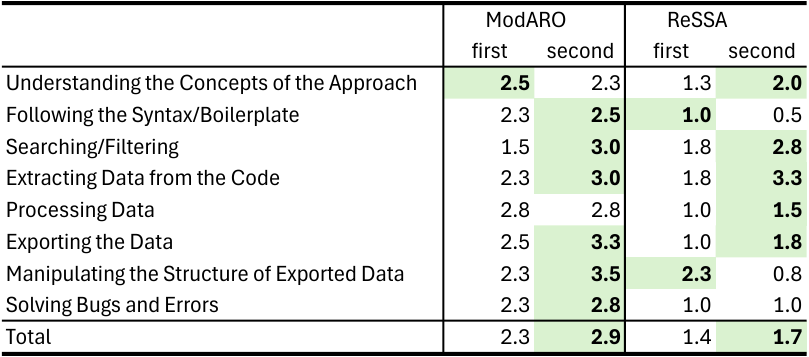}
    \caption{
        Comparison of mean challenge responses broken down by tool usage order.
        The better result for each approach across the two
        groups is highlighted in green.
    }
    \label{tab:challange-groups}
\end{figure}

As shown in Figures~\ref{fig:challenges2} and~\ref{fig:participants-matrix},
participants rated ModARO as less challenging or equally challenging as ReSSA for
all activities on average, reflecting the responses to the ease-of-use questions.
When sorted according to which technology was used first,
there appears to be a small difference in responses
between the two participant groups: participants in group 1 (ModARO first)
reported finding ReSSA easier to use than those in group 2 (ReSSA first).
We explore this in Figure~\ref{tab:challange-groups}, which shows that a tool tended to
be reported as easier by whichever group used it second.
This suggests that the order in which the tools were used is a source of bias,
but the sample sizes are too small to establish statistical significance.
Such a bias may also have contributed to the positive results
for the Hybrid approach, which was always used third.

It was unanimously reported that our approach was easier when it came
to manipulating the structure of exported data and when solving bugs
and errors.
There was also a near unanimous preference for our approach with regard to the ease of
syntax/boilerplate, data processing, and data export activities. 
Participant D3, the only participant who rated ReSSA better at
data processing and exporting, reported elsewhere in the questionnaire
that they had great trouble
with Regex, which may explain their response.
Participant D5 found the syntax and boilerplate 
to be more challenging in our approach and later referenced
confusion around how the inputs and outputs of an extractor
are defined.
The activities for which our approach received poorer results
were searching, filtering, and extracting data from code.
In these two activities, all the participants in group 1 found
our approach to be more or equally challenging as ReSSA, while for
group 2, it was the reverse. Taken together, participants
appeared to find both approaches equally challenging when performing
these activities.

\subsubsection{Thematic Analysis of Feedback}
The three most common themes were as follows:

\textit{Coding and Tooling Over Config:}
The most common theme among participants was the preference for
code over the configuration and string callback approach used by ReSSA.
When asked what they liked about ModARO,
five of the eight participants made statements such as ``it was all code'',
``it is much more intuitive as it requires coding rather than config'',
and ``Javascript syntax and logic'', while a sixth stated they liked being
``able to process data in real code, rather than a string containing code''.
Participants showed a particular dislike for the ReSSA single-line string
callbacks, saying that it was ``exceptionally challenging...
near impossible to comprehend a simple loop'', ``very painful to understand'',
and that ``the logic is great but the formatting is horrible, it's not human at all''.
The contrast between the
ReSSA callbacks and our approach was likely exacerbated
by the language tooling provided by the participants'
code editor (Visual Studio Code),
which provided syntax highlighting, code completion,
and type hints for the ModARO API.
At least two participants resorted to writing the Rune
code in separate files first so that they
could use indentation, multiple lines, and syntax highlighting
to avoid simple syntax errors.

Most of the criticisms of ReSSA relate specifically to
the single-line string callbacks
and not the declarative patterns for matching
abstract code structures. This is reflected by the positive 
feedback for the hybrid approach, which used the same
pattern matching as ReSSA but
was free of most of the Rune coding as our reconstruction algorithm handled
identifying the microservices and ran the analysers
on each individually, hence entirely avoiding
the need for most code statements and structures
(for loops, if statements, etc).
It is important to recognise that this is not a question
of declarative vs imperative extraction; both ModARO and
ReSSA have imperative (extractor functions vs string callbacks)
and declarative (regex vs ReSSA patterns) elements. This feedback
is instead a reflection of how well those elements were integrated
and the tooling support to facilitate their use; participants found it easier
to write code and define declarative structures for ModARO than to
write the JSON files and string callbacks for ReSSA.
These answers likely explain or contribute to why our approach was rated
easier to use than ReSSA, and why our approach was seen as less challenging
for most of the challenge questions.

\textit{Better Parsing Capabilities:}
The next most common theme in participant responses was the
desire for better parsing capabilities. The current focus on regular
expressions was the most disliked aspect of ModARO;
participants stated that the ``regex got very complex,
simplifying that would make [ModARO] more appealing'', and that
``complex regex makes it confusing to extract the data in an easy way''.
One participant directly stated that ``if there was a way to simplify
the interaction with regex [they] think everything else is straightforward''.
Several participants brought up integrating the structured parsing
ReSSA provides into our approach,
but not at the expense of the elements brought up in the previous theme.
One participant said that they were ``curious for a code version of ReSSA.
The ReSSA filtering was good. Somehow adding that to the
code would be nice'', while another
said that the hybrid approach specifically was a
``huge improvement'' when it came to parsing code files.
One suggested that some of the simple API functions we provided to search files
could be modified to accept strings instead of regular expressions when
performing very simple substring searches.
These responses likely explain why our approach was rated similarly challenging
to ReSSA for data filtering and extraction tasks.

Based on this feedback, it appears that the core of our
approach, the modular extractor scripts and reconstruction algorithm, is
broadly popular among the participants but that to be more
widely useful, we would need to provide APIs in the code to handle
parsing at a higher level of abstraction.
The responses to the hybrid approach in
Figure~\ref{fig:usefulness} show how participants reacted positively to an
approach with more advanced parsing, but with simplified
data manipulation due 
to our recursive algorithm workflow.

\textit{The Steep Learning Curve:} \label{section:theme/learning-curve}
Half of the participants made note of the steep learning curve
during the study; unsurprising given that we were asking them to
learn, comprehend, and use
several tools in a relatively short time frame.
Participants stated that they found the ``learning curve'' challenging
and that ``cognitive overload may skew usability results''.
When asked what would make ModARO easier to use, three
responded ``familiarity with the tool'', ``more time'', and
``more experience using the tool -- would decrease mental effort''.
These responses show optimism that the approach and tool would become
easier to use over time and implies that the ease-of-use responses
were negatively biased due to the short duration of the study. This bias would
presumably apply to both tools, though it is possible that extended use would
show one tool to be more intuitive than the other in a way not represented
within a short timeframe.

\begin{tcolorbox}[rq]
    \textbf{RQ4 Answer:}
        In general, ModARO was considered more useful and usable than the baseline. Participants found it easier
        to manipulate data, debug errors, and understand conceptually, and cited the
        code-based workflow, enhanced by syntax highlighting and type hints,
        as a key usability advantage. However, both approaches were viewed as
        similarly challenging in filtering and extracting data from code,
        with participants expressing a desire for more advanced code-parsing
        tools than Regex.
\end{tcolorbox}

\section{Limitations \& Threats to Validity}
\subsection{Case Study}

The identification of architectural information documented by different projects
and the categorisations of why information could or could not be extracted
were both the work of a single researcher, which increases the risk that
documented information was missed or that information we could not collect
was actually present in the codebase.
However, this information is not vital for answering the
research question; we only need to collect enough information to
demonstrate that our approach supports diverse codebase.
Missing information is only an issue if that information reveals
a limitation of our approach, and given the lack of such evidence
from the considerable information we did collect, there is no
indication that this would be the case for any information missed.
In addition, the extractors written for this study were all written by the same
researcher and it would be likely that other individuals would have
written them differently, potentially consolidating multiple extractors
into one or vice versa. For this reason, we focus on the
reasons that extractors were not reusable instead of the absolute number.
It is also possible that we made mistakes in our extractor categorisations,
as only 15\% were verified by two reviewers.
However, only one miscategorisation was identified, and it was due to the first
reviewer failing to recognise a technology; further miscategorisations
of this type would only strengthen our results by showing that more
extractors are reusable.

\subsection{User Study}
In the user study, we wanted to evaluate the usefulness and ease-of-use
of our approach, as well as collect general feedback by testing our
approach with industry practitioners.
However, we only tested the approach with eight people 
and they all worked in the same company, so our results may not be
broadly representative of industry practitioners.
To mitigate this issue we tried to recruit participants with different experience levels 
and who worked on different projects to broaden our results.
In addition, the participants only had a short time with the tools
and the tasks they were asked to perform were necessarily simple due to 
time constraints. If participants had been able to use the tools for days
or weeks and given more complex tasks, the results may have been different,
but the participants' time was too limited for such a study to be feasible.
Finally, the way the sessions were run may have influenced the results;
participants were taught the tools in groups of four, and so questions
asked by an individual would have influenced the rest of the group.
In addition, the teaching material itself may have done a better job of
explaining one tool over another.
We tried to mitigate these issues by conducting numerous pilot studies
to refine the teaching material and identify points of confusion that
pilot testers asked about that could be directly addressed by the teaching material.

\section{Discussions, Implications, \& Future Work}
Our results show that our approach can successfully reconstruct
microservice architectures built with diverse programming languages,
libraries, and deployment technologies.
Modularity was a success, with the majority of the reusable
extractors created during the study not requiring specific
modifications to work across projects. The only exceptions, the Docker
Compose variant extractors, would not have been necessary if the
extractors were configurable; a minor change which we
implemented between the studies.
A significant finding was that not all extractors can be reusable; some information is
inherently project-specific, meaning that many projects will still require custom code.
However, our approach allows project-specific code to be seamlessly
used alongside the reusable extractors without the latter requiring modifications,
which seems the best-case scenario as it minimises the
code that must be rewritten for each project.
That said, there is still work to be done to realise the full benefits
of reusability.
Firstly, a shared repository of extractors for common technologies would
be of great benefit, reducing the need for individual users or projects
to create their own.
In addition, for these extractors to be reusable, it is essential that
a shared architecture model schema be agreed upon to store the extracted information;
without a shared model, there may be conflicts,
extractors will not be able to inherently pass
information to each other, and results cannot be successfully aggregated.
While we used a simple schema for testing, that schema was not
based on any research and is unlikely to be sufficient for every project.

We also did not explore the questions around parsing code files (using proper syntax parsers
as opposed to regular expressions):
how best to provide this capability to extractors, how frequently such parsing
is needed (compared with simply reading configuration files and
file paths), and which kinds of architectural information would most likely
benefit from such parsing.
Participants in the user study were eager for
such functionality, but we deliberately gave them problems that required
code analysis to allow a better comparison with ReSSA. The significance of that feedback would
depend on how often code analysis is required in practice.

The features designed to
support reconstruction of distributed codebases were also effective;
extractors were able to be reused between mono- and multi-repo projects
without changes, and the aggregation and retroactive linking were successful
for the two multi-repo study projects. However, we did find that the
link schema functionality described in Section~\ref{section:design/links}
was insufficient; for some project configurations (e.g., domains for HTTP requests dynamically
set in remote repositories),
link schemas must not only specify and restrict field types and values,
but also allow these types and values to be retroactively resolvable.
This presents an engineering challenge that still needs to be overcome,
but it does not fundamentally change the nature of our approach.
This scenario is also a good example of the challenges inherent
to static architecture reconstruction
of distributed systems.
The next step to evaluate the concept of distributed architecture reconstruction is to
integrate the tooling into
the CI/CD pipelines of real multi-repo projects, but there are still some
unexplored questions; specifically, how best to collect model files for
multiple microservices after they have been individually created,
and how usable and useful is such a workflow in practice?

\section{Related Work}
Existing automated reconstruction approaches for microservice
architectures can broadly be divided into two categories;
static analysis, which uses artefacts such as source code and configuration files;
and dynamic analysis, which typically uses data from network traffic,
log outputs, and trace data~\cite{Cerny2022a}.
Existing research has often focused on dynamic analysis~\cite{Engel2018,Mayer2018,Kleehaus2018},
or a hybrid approach that extracts some metadata
from the code repository but gathers inter-service
communication information dynamically~\cite{Granchelli2017,Ma2018}.
However, dynamic analysis requires the computational resources to run the
full architecture and enough time to observe microservice behaviours,
making it unsuitable for resource-constrained CI/CD pipelines,
especially in multi-repo projects where each pipeline sees only one microservice;
hence our focus on static analysis.
Much static analysis research requires the usage of specific technologies,
such as Docker and Docker Compose~\cite{Ibrahim2019},
gRPC calls~\cite{Fang2023},
Kubernetes~\cite{Muntoni2021,Soldani2021,Soldani2023},
or Java Spring~\cite{Bushong2021,Ma2018}. These approaches are not
analogous to ours, which is technology-agnostic.

Ntentos et al.~\cite{Ntentos2021}
propose an approach with programmable 'detectors'; class-based entities
for parsing and validating the static artifacts of an application
which developers can write and re-use between projects to
support different technologies.
However, the focus of this approach is architecture verification, not reconstruction;
users must explicitly define the entities to search for, and must manually pass
data to and between detectors.
There is also the ReSSA approach~\cite{Schiewe2022a},
which we used for our user study.
The Language Agnostic Abstract Syntax Tree (LAAST) used by this
approach theoretically allows support for any language, but it presents
a significant barrier of entry when adding support for new languages,
which requires the identification of any and all
language-specific structures and statements (ternaries, match/switch, goto, etc.)
and the conversion of them into equivalent LAAST constructs,
or the modification of the LAAST if a language's features are impossible
to translate~\cite{Bushong2022a}.
ReSSA's focus is on making it easier to analyse code in
multiple languages; modularisation of that analysis code, the main contribution
of our work, is not a focus, and their approach is unable to collect information
from non-code files.
Schneider and Scandariato~\cite{schneider2023} present another
language agnostic static analysis technique. They similarly divide
their reconstruction code into technology-specific modules (coincidentally also
called `extractors')
and propose using a shared data flow diagram
to pass data between extractors in a similar manner to our
architecture model.
This has similarities to our approach, but the focus of the
research is on the methods of extracting data from static files
and measuring the accuracy of the results. The research lacks our
investigation of modularity, which is demonstrated by the lack of an
algorithm to automatically determine the execution order of extractors.
In addition, they do not consider the challenges of reconstructing
multi-repository projects, nor does any other research that we are aware of.

\section{Conclusion}
In this paper, we presented and evaluated ModARO, an approach to
microservice architecture reconstruction that enables the development
of modular reconstruction code (`extractors') for any technology and
their reuse across different projects, independent of the surrounding
technology stack or whether an extractor operates on mono- or multi-repo
projects.
We found that our approach was successful in extracting many types
of architectural information from a range of different technologies
and that practitioners found it both useful and easy to use.
The extractors proved to be reusable between projects, although some information
was inherently project-specific.
Distributed reconstruction was also a success; extractors were reusable in both
mono- and multi-repo contexts, and we were able to specify
links between repositories,
although some project configurations require more complex link functionality.
The findings allowed us to identify improvements that could be made in both areas to better
facilitate modularity and distributed reconstruction.
This approach allows developers to assemble or create extractors tailored to
their technology stack and distribute architecture reconstruction
across microservice repositories, enabling integration into repository CI/CD
pipelines and hence facilitate continuous documentation,
architecture conformance validation, or change detection workflows.

\section*{Acknowledgements}
We thank Swordfish Computing for allowing their employees to
participate in the user study and for partially funding
this research. We also thank Aaron Prince for
assisting with participant recruitment, and we thank
Lachlan Maddaford,      %
Saba Sadeghi Ahouei,    %
Joshua Francis,         %
and Liam Wigney         %
for participating in pilot studies.

\bibliographystyle{elsarticle-num}

\end{sloppy}

\end{document}